\PassOptionsToPackage{hidelinks}{hyperref}
\documentclass[sigplan,11pt,nonacm=true,anonymous=false]{acmart}
\usepackage{algorithmic}
\usepackage{graphicx}
\usepackage{textcomp}
\usepackage{xcolor}
\usepackage{array}
\usepackage{siunitx}
\usepackage{makecell}
\usepackage{tikz}
\usepackage{fontawesome}
\usepackage{multirow}
\usepackage{xfrac}
\expandafter\def\expandafter\UrlBreaks\expandafter{\UrlBreaks%
	\do\a\do\b\do\c\do\d\do\e\do\f\do\g\do\h\do\i\do\j%
	\do\k\do\l\do\m\do\n\do\o\do\p\do\q\do\r\do\s\do\t%
	\do\u\do\v\do\w\do\x\do\y\do\z\do\A\do\B\do\C\do\D%
	\do\E\do\F\do\G\do\H\do\I\do\J\do\K\do\L\do\M\do\N%
	\do\O\do\P\do\Q\do\R\do\S\do\T\do\U\do\V\do\W\do\X%
	\do\Y\do\Z\do\*\do\-\do\~\do\'\do\"\do\-}%

\usetikzlibrary{arrows.meta}

\makeatletter
\def\@centerlast{\leftskip=\z@ \@plus .5fil\relax \rightskip=\z@ \@plus -.5fil\relax
\parfillskip=\z@ \@plus 1fil\relax}

\def\capaddon#1{\medskip\par\parbox\linewidth{%
\@centerlast\itshape\scriptsize#1}\par}

\usepackage{listings}
\usepackage{multicol}
\usepackage{forest}
\usetikzlibrary{tikzmark}
\def\HStrut{\vphantom{\{tg}}%
\def\Snode#1{\tikzmarknode{#1}\HStrut}%

\useforestlibrary{edges}

\usepackage[super]{nth}
\usepackage[inline]{enumitem}
\newlist{inlist}{enumerate*}{1}
\setlist[inlist]{itemjoin={{, }},itemjoin*={{, and }},label=($\roman*$),mode=boxed}
\newlist{orlist}{enumerate*}{1}
\setlist[orlist]{itemjoin={{, }},itemjoin*={{, or }},label=($\alph*$),mode=boxed}
\newlist{andlist}{enumerate*}{1}
\setlist[andlist]{itemjoin={{, }},itemjoin*={{, and }},label=($\alph*$),mode=boxed}

\long\def\AnswerBox#1#2{\smallskip\par\noindent\fbox{\parbox{\dimexpr\linewidth-2\fboxsep-2\fboxrule}{\textbf{Answer to RQ#1}:~\textit{#2}}}\medskip\par}
\def\zcline{\expandafter\expandafter\expandafter\zzcline\cline}
\def\zzcline#1\cr{#1\smash{~\;}\cr}
\def\xcline{\expandafter\expandafter\expandafter\xzcline\cline}
\def\xzcline#1\cr{#1\smash{~~~\;}\cr}

\usepackage{xspace}
\def\FMeasure{F\textsubscript{1}\xspace}

\usepackage[capitalise]{cleveref}
\usepackage{csquotes}
\let\say\enquote
\let\T\texttt
\usepackage{parcolumns}
\usepackage{float}
\usepackage{flushend} %
\usepackage{orcidlink} %

\let \tldots = \ldots
\def \ldots {\textrm{\tldots}}

\forestset{
   depthWildcardL/.style={edge label={node[midway,left,darkgray] {\footnotesize\ttfamily\(\downarrow\)\,*}}},
   depthWildcardR/.style={edge label={node[midway,right,darkgray] {\footnotesize\ttfamily*\,\(\downarrow\)}}},
   optional/.style={edge label={node[midway,right,darkgray,yshift=1mm] {\footnotesize\ttfamily?}}},
   subtree/.style={for tree={font=\ttfamily,align=center,s sep+=1em,l sep+=.25em,edge={dashed,gray,very thick}}},
   subtree tight/.style={for tree={font=\ttfamily,align=center,s sep=0em,l sep+=.25em,edge={dashed,gray,very thick}}}
}
\def\WithMod#1#2{#1\\[-.8ex]\scriptsize\textit{#2}}
\def\AsAny#1{\WithMod{#1}{UNORDERED}}%
\def\BindTo#1{(\textcolor{black}{\textit{#1}})}

\tikzset{matchextraopts/.style={}}
\def\Match(#1)#2;{\node[fill=black!90!gray,circle,inner sep=1.5pt,text=white,font=\bfseries\sffamily\small,below right,opacity=.9,minimum size=3ex,scale=.72,matchextraopts] at (#1) {\clap{#2}};}

\newcommand{\mycomment}[1]{}

\tikzset{blob/.style={draw=black,circle,minimum size=3mm},K/.style={black,scale=.5,xscale=.75}}
\newsavebox\minisubtreeA
\setbox\minisubtreeA=\hbox{\begin{tikzpicture}[K]
	 \node[blob] (@) at (0,0) {};
	 \node[blob] (@1) at (-1,-1) {};
	 \node[blob] (@2) at (1,-1) {};
	 \draw (@1) -- (@) -- (@2);
\end{tikzpicture}}
\newsavebox\minisubtreeAb
\setbox\minisubtreeAb=\hbox{\begin{tikzpicture}[K]
	 \node[blob] (@) at (0,0) {};
	 \node[blob,fill] (@1) at (-1,-1) {};
	 \node[blob] (@2) at (1,-1) {};
	 \draw (@1) -- (@) -- (@2);
\end{tikzpicture}}
\newsavebox\miniast
\setbox\miniast=\hbox{\begin{tikzpicture}[K]
	 \node[blob] (@) at (0,0) {};
	 \node[blob] (@1) at (-1,-1) {};
	 \node[blob,fill] (@2) at (1,-1) {};
	 \node[blob] (@4) at (0.5,-2) {};
	 \draw (@1) -- (@) -- (@2) -- (@4);
\end{tikzpicture}}
\newsavebox\minisubtreeB
\setbox\minisubtreeB=\hbox{\begin{tikzpicture}[K]
	 \node[blob] (@) at (0,0) {};
	 \node[blob] (@1) at (-1,-1) {};
	 \node[blob] (@2) at (1,-1) {};
	 \node[blob] (@3) at (1.5,-2) {};
	 \node[blob] (@4) at (0.5,-2) {};

\draw (@1) -- (@) -- (@2) (@4) -- (@2) -- (@3);
\end{tikzpicture}}
\newsavebox\minisubtrees
\setbox\minisubtrees=\hbox{\begin{tikzpicture}[black]
	 \node (@) at (0,0) {\usebox\minisubtreeAb};

\draw[Kite-Kite] ([xshift=1mm]@.east) to[edge node={node[above] {\sffamily?}}] ++(5mm,0);
	 \node[right] (@2) at ([xshift=7mm]@.east) {\usebox\miniast};
\end{tikzpicture}}

\def\BibTeX{{\rm B\kern-.05em{\sc i\kern-.025em b}\kern-.08em
T\kern-.1667em\lower.7ex\hbox{E}\kern-.125emX}}
\def\todo#1{\textsf{\textcolor{purple}{\textbf{TODO:}} #1}}
\def\todo#1{}%
\def\who#1{\textsf{\textcolor{blue}{\textbf{#1}}}}
\def\who#1{}

\newcommand\copyrighttext{%
  \footnotesize \textcopyright 2024 IEEE. Personal use of this material is permitted.
  Permission from IEEE must be obtained for all other uses, in any current or future
  media, including reprinting/republishing this material for advertising or promotional
  purposes, creating new collective works, for resale or redistribution to servers or
  lists, or reuse of any copyrighted component of this work in other works.
  DOI: \href{https://doi.org/10.1109/ICCQ60895.2024.10576984}{10.1109/ICCQ60895.2024.10576984}}
\newcommand\copyrightnotice{%
\begin{tikzpicture}[remember picture,overlay]
\node[anchor=south,yshift=10pt] at (current page.south) {\fbox{\parbox{\dimexpr\textwidth-\fboxsep-\fboxrule\relax}{\copyrighttext}}};
\end{tikzpicture}%
}

\begin{document}

\title{Exploring the Effectiveness of Abstract Syntax Tree Patterns for Algorithm Recognition}

\author{Denis Neumüller\,\orcidlink{0000-0003-3872-0188}} 
\affiliation{%
  \institution{Ulm University}
  \country{Germany}
}
\email{denis.neumueller@uni-ulm.de}

\author{Florian Sihler\,\orcidlink{0000-0001-7195-7801}}
\affiliation{%
  \institution{Ulm University}
  \country{Germany}
}
\email{florian.sihler@uni-ulm.de}

\author{Raphael Straub\,\orcidlink{0009-0007-8534-0053}}
\affiliation{%
  \institution{Ulm University}
  \country{Germany}
}
\email{raphael.straub@uni-ulm.de}

\author{Matthias Tichy\,\orcidlink{0000-0002-9067-3748}}
\affiliation{%
  \institution{Ulm University}
  \country{Germany}
}
\email{matthias.tichy@uni-ulm.de}

\begin{abstract}
The automated recognition of algorithm implementations can support many software maintenance and re-engineering activities by providing knowledge about the concerns present in the code base.
Moreover, recognizing inefficient algorithms like Bubble Sort and suggesting superior alternatives from a library can help in assessing and improving the quality of a system.
Approaches from related work suffer from usability as well as scalability issues and their accuracy is not evaluated.
In this paper, we investigate how well our approach based on the abstract syntax tree of a program performs for automatic algorithm recognition.
To this end, we have implemented a prototype consisting of: A domain-specific language designed to capture the key features of an algorithm and used to express a search pattern on the abstract syntax tree,
a matching algorithm to find these features, and an initial catalog of \say{ready to use} patterns.
To create our search patterns we performed a web search using the algorithm name and described key features of the found reference implementations with our domain-specific language.
We evaluate our prototype on a subset of the BigCloneEval benchmark containing algorithms like Fibonacci, Bubble Sort, and Binary Search.
We achieve an average \FMeasure-score of 0.74 outperforming the large language model Codellama which attains 0.35.
Additionally, we use multiple code clone detection tools as a baseline for comparison, achieving a recall of 0.62 while the best-performing tool reaches 0.20.

\end{abstract}

\maketitle
\copyrightnotice

\vspace{-2em}
\paragraph{Keywords: algorithm recognition, program comprehension, pattern matching, abstract syntax tree, domain-specific language, reverse engineering, maintenance} %
\section{Introduction} %
Program comprehension is one of the major activities developers conduct during software development.
Empirical research shows that developers spend more than half of their time on this activity~\cite{DBLP:journals/tse/XiaBLXHL18}. %
One area of interest in program comprehension is the recognition of design decisions, e.g., the recovery of models~\cite{DBLP:journals/access/RaibuletFZ17}, the recovery of design-patterns~\cite{DBLP:journals/air/YarahmadiH20}, the recovery of software architecture design decisions~\cite{DBLP:conf/icsa/ShahbazianLLBM18}.

While those approaches aim at understanding the software's big picture, we are interested in design decisions on a smaller scale, i.e., recognizing which algorithms have been used in implementing the software's functionality.
We follow the definition by Cormen et al.~\cite[p. 5]{DBLP:books/daglib/0023376} that an algorithm \say{describes a specific computational procedure~[\ldots] for solving a well-specified computational problem}, e.g., sorting, handling data structures, computing shortest paths in a graph, scheduling processes, achieving consensus in a distributed system.

Multiple algorithms with distinct characteristics exist for solving the same computational problem.
For example, Cormen et al.~\cite[p. 148ff.]{DBLP:books/daglib/0023376} give a nice overview of the differing characteristics of multiple sorting algorithms, e.g.,~average and worst case time complexity, sorting in-place or not, whether the probabilistic distribution of the data is known.
Accordingly, Singh and Sarangi \cite{DBLP:conf/msr/SinghS20} show that different data structures and algorithms implemented in different operating systems for the same functionality result in real-life performance differences.
Consequently, the information of which algorithm has been implemented in a system gives a developer insights into the design decisions of the original developers and the system's quality.
Correspondingly, Anquetil et al.~\cite{DBLP:journals/infsof/AnquetilOSD07} and Rashid et al.~\cite{DBLP:journals/ijinfoman/RashidCO19} state that knowledge about which particular algorithms are used in a software system is required for its maintenance.

Determining whether two programs are semantically equivalent — and therefore algorithm recognition — is impossible in the general case~\cite{Metzger2000}.
This means that all methods to tackle the problem, including ours, use a best-effort approach.
Related works such as Mesnard et al.~\cite{Mesnard2016} and Metzger et al.~\cite{Metzger2000} try to recognize algorithms by formally proving the equality of two pieces of code.
Due to the undecidability, these approaches only work within certain limitations e.g. only being able to analyze a subset of well-formed programs or providing inconclusive results.
Using machine learning for the classification of algorithms can often achieve good results on the respective test sets~\cite{Shalaby2017, Long2022, Zhu2011, Taherkhani2010}.
However, these approaches do usually not address how the search for an algorithm implementation consisting of more than a single method would take place on a real-world code base.
Furthermore, extending these approaches to algorithms outside the training set is not easily possible since this requires obtaining a suitable dataset and retraining the models.
The area of program concept recognition focuses on the recognition of high-level concepts including algorithms, data structures, and higher-level abstractions~\cite{Kozaczynski1992, Quilici1994, Wills1994, Nunez2017}.
These approaches often use multiple complementary analysis techniques like abstract syntax trees (ASTs), program dependency graphs (PDGs), or data-flow graphs in combination with control-flow or data-flow constraints. However, none of those approaches evaluate the recognition performance of their search patterns.

Our aim in this paper is to explore how well algorithms can be recognized by algorithm search patterns simply based on the AST without extra analysis like computing PDGs.
We focus on this approach because we believe that specifying search patterns in terms of AST structures is akin to writing code and therefore easier for developers than having to think solely in terms of data- and control-flow dependencies.
We support automatically injecting information about the recognized algorithms as comments into the source code at the algorithms' implementation location. 
This can be done either as a one-time activity for the full code base or incrementally as part of the continuous integration pipeline.
We therefore focus on detecting specific algorithm implementations such as Binary Search or Bubble Sort rather than attempting to generalize our search patterns across multiple different search or sorting algorithms.

Another envisioned use case for our approach is supporting the quality assurance of applications.
Detecting subpar algorithm implementations in a code base and suggesting the replacement with a better-suited alternative from a high-quality library allows one to access and enhance the code quality.
Similar to other static code analysis tools this activity could take place as part of a project's build pipeline.

The contributions of this paper are
\begin{inlist}
	\item a language for specifying search patterns on abstract syntax trees
	\item an algorithm finding all matches of those patterns in code
	\item a set of example pattern specifications for selected algorithms
	\item the evaluation of those on BigCloneEval
\end{inlist}. %

Our evaluation of six different algorithms shows that it is indeed possible to recognize algorithms just by specifying search patterns on the abstract syntax tree.
We were able to achieve good precision and recall for all algorithms but Binary Search, which is lacking in recall.

We then compare ourselves with Codellama~\cite{roziere2024codellama} a large language model which has shown impressive results on other code-related tasks~\cite{roziere2024codellama, liu2024LLMCodeGenEval, silva2023repairllama}.
Codellama achieves similar recall but has significantly worse precision leading to a macro-averaged \FMeasure-score\footnote{${Macro-average\;\: \FMeasure-score} = \frac{F_{1_{Fibonacci}}\: +\: \hdots\: +\: F_{1_{Binary Search}}}{6}$} of 0.35 while we achieve 0.74.
In addition, the LLM has a significantly longer runtime orders of magnitude above ours.

Afterwards, we compare our approach to the code clone detection (CCD) tools CloneWorks, SourcererCC, NiCad, and Oreo.
We consider CCD tools as a baseline since they can also be used to recognize algorithms by executing them with one or several algorithm reference implementations as input.
We outperform the CCD tools for all six algorithms, achieving a macro-averaged recall of 0.62 with the best CCD tool achieving 0.20.
Particularly, our approach yields a significantly higher recall for Type 3 and Type 4 code clones.

The following section introduces our abstract syntax tree-based pattern language and the corresponding matching procedure on source code.
\Cref*{sec-evaluation} presents the results of our evaluation followed by a comparison with Codellama and finally a comparison with CCD tools as a baseline.
\Cref*{sec-relatedwork} discusses different approaches from the related work in comparison to ours.
\Cref{sec-conclusion} concludes and gives an outlook on future work.

\section{A Pattern Language for Algorithm Recognition}\label{sec-prototype}
\begin{figure}[H]
	\centering
	\includegraphics[width=0.48\textwidth]{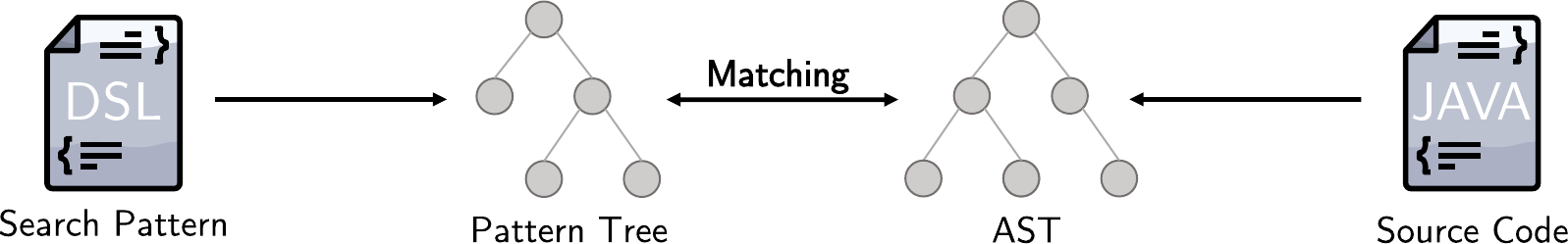}
	\caption{Overview of our approach.}
	\label{fig:overview}
\end{figure}
\cref{fig:overview} gives an overview of our approach.
The first step in our framework is the specification of a search pattern in our domain-specific language (DSL) which is embedded in Java.
The DSL is designed to describe parts of the AST that the user considers to be key features of an algorithm.
In the next step, the search pattern is transformed into a low-level pattern tree, that will be matched to the AST during the matching process.
Additionally, the transformation serves the purpose of translating syntactic sugar and convenience methods, which form an integral part of the DSL, into the basic language primitives of our approach.
We then use the source code analysis and transformation framework Spoon~\cite{Spoon} to parse the Java files in the given code base into their respective ASTs.
Finally, we start the matching process in which we compare the structure and values of the pattern tree with each of the ASTs.

A language for describing algorithms must be specific enough to describe any algorithm in sufficient precision so only relevant instances are recognized as such.
At the same time, the language must be general enough to abstract from implementation details.
We designed our internal Java DSL to make it intuitive to describe (only) the parts of the AST that the user considers to be key features of the algorithm.
The language allows to specify the permissiveness of a search pattern, e.g., by abstracting from details of the AST.
While an exact description of the source code AST would also be a valid search pattern, abstracting details of the AST is an important feature to find different instances of an algorithm despite implementation variances.
The overall design of our DSL is heavily inspired by the fluent interface style promoted by Fowler et al. in their book Domain-Specific Languages~\cite{FowlerDSLs2010}.
To create our DSL, we analyzed multiple algorithm implementations to identify their key features.
This shaped the core primitives of the language and culminated in an early version of the DSL.
We then used this version to describe various algorithms, through which we identified improvement opportunities such as the convenience features. %
Afterward, we incorporated the identified improvements resulting in the DSL in its current form.
In the following, we first describe the basic primitives needed to describe search patterns before going on to discuss convenience methods that increase usability and facilitate abstraction in \cref{sec-convenience}.

\subsection{Core Language Primitives}\label{core-languge-primitives}
\begin{figure*}[t]
\noindent\begin{minipage}[t]{.45\textwidth}
	\begin{lstlisting}[caption=Basic prime factors search pattern,label=list:search-pattern-basic,basicstyle=\scriptsize\ttfamily,language=Java,escapechar=|, numbers=left, numberstyle=\tiny, numbersep=2pt, xleftmargin=7pt]{Name}
method().bindTo("method").body(block()|\label{l1:method}||\label{l1:bindmethod}|
 .next(wideWildcard()) |\label{l1:wildcard1}|
 .next(forLoop()|\label{l1:forLoop}|
				 .condition(binOp().ops("<", "<="))|\label{l1:condition}|
				 .body(block().next(
						whileLoop()|\label{l1:whileLoop}|
							.condition(binOp().ops("%", "/") |\label{l1:conditionWhile}|
									.lhs(varRead().bindTo("num")) |\label{l1:bindnum}|
									.rhs(varRead().bindTo("index")) |\label{l1:bindindex}|
							)
							.body(block()
									.next(wideWildcard())
									.next(assignment() |\label{l1:assStart}|
											.lhs(varWrite().bindTo("num"))|\label{l1:bindusenum}|
											.rhs(depthWildcard(varRead().bindTo("index")))) |\label{l1:assEnd}||\label{l1:binduseindex}||\label{l1:depthWildcard}|
									.next(wideWildcard())))))
 .next(wideWildcard())
 .next(optional(returns())) |\label{l1:ret}|
	\end{lstlisting}
	\end{minipage}\hfill
	\noindent\begin{minipage}[t]{.45\textwidth}
	\begin{lstlisting}[caption=Example prime factors implementation,basicstyle=\scriptsize,language=Java,escapechar=|,label=list:primefactors-impl, numbers=left, numberstyle=\tiny, numbersep=2pt]{Name}
List<Integer> getPrimeFactors(int n) {
	var primes = new LinkedList<Integer>();|\label{lpf:listdef}|
	for (int i = 2; i <= n; i++) {
			while (n % i == 0) {
					primes.add(i);|\label{lpf:listwrite}|
					n /= i;
			}
	}
	return primes;
}
	\end{lstlisting}
	\end{minipage}
	\caption{A Search Pattern for Prime Factors using only core language primitives.}
	\label{fig:primefactors-core}
\end{figure*}

\cref{fig:primefactors-core} illustrates a search pattern for the prime-factors algorithm expressed using only the core primitives of our language.
With the pattern, we search for the following:
\say{We expect a method binding the method name to the specified identifier, in this case, \enquote*{method} (\Cref{l1:method}).
We then specify the statements contained in the method body with \T{block.\allowbreak next(stmt)}. %
Using the \T{wide\-Wildcard()} at multiple places in the search pattern allows us to ignore statements that are not central to the logic of the prime factors calculation
e.g. the list definition in \cref{lpf:listdef} or unrelated code such as logging.
Among the statements inside the method body, we expect a for-loop that uses a comparison with \(<\) or \(\leq\) as the condition (\crefrange{l1:forLoop}{l1:condition}).
The body of the for-loop should contain a while loop that uses a comparison with \(\%\) or~\(/\) as the condition (\crefrange{l1:whileLoop}{l1:conditionWhile}).
We bind the left-hand side of the binary operation to the identifier \enquote*{num} and the right-hand side to the identifier \enquote*{index} (\crefrange{l1:bindnum}{l1:bindindex}).
Among the statements contained inside the while loop, we expect an assignment that writes the variable previously bound by \enquote*{num} with the \T{depth\-Wildcard()} specifying that the variable \enquote*{index} must be present somewhere inside the right-hand side of the assignment (\crefrange{l1:assStart}{l1:assEnd}).
At the end of the method, we allow an optional return statement to be present (\Cref{l1:ret}).}
The features displayed in the example can be divided into the following basic core language primitives:

\subsubsection{Builders}
 Our DSL exposes builder classes for matching individual Java language constructs. Examples include builders for binary operations (\T{bin\-Op()}), assignments~(\T{as\-sign\-ment()}), for-loops~(\T{for\-Loop()}) or return statements~(\T{re\-turns()}).
 The builders expose construct-specific configuration options to the user such as the operator kind for \T{bin\-Op()} or the condition for \T{for\-Loop()}.
 In case the user does not provide a configuration, our language is permissive by default meaning that we would match a binary operation with any operand or a for-loop with any condition.
 Additionally, the \T{any()} and \T{any\-Type()} constructs allow the user to match any AST element or any type reference in the AST, similar to the \enquote{.} in regular expressions.
 When specifying a search pattern, the builders are used as basic building blocks, whose combination then results in a meaningful pattern, e.g., \T{bin\-Op().\allowbreak rhs(var\-Read())} matches a binary operation with a variable read on the right-hand side.

\subsubsection{Binding Constraints}
 We allow binding match\-ed parts of the AST to a user-defined identifier by using the \T{bindTo(id)} construct.
 This identifier can be used multiple times in the search pattern to require that parts of the AST reference the same element.
 The provided example uses this to make sure that the same variables are used in the condition (\crefrange{l1:bindnum}{l1:bindindex}) and body (\crefrange{l1:bindusenum}{l1:binduseindex}) of the while loop.
 Binding constraints can also be used to specify that a method calls itself recursively by binding the signature of the method definition (as shown \Cref{l1:bindmethod}) and requiring a \T{method\-Call()} in the body, again binding the signature to the same identifier.

In addition, binding constraints can be used to link different search patterns.
 For example, when searching for implementations of the quicksort algorithm, separate search patterns for the partition and swap implementations can be created.
 By binding the signature of the swap implementation, it is now possible to require that the swap is called in the partition search pattern.

\subsubsection{Wildcards}
 Sometimes it is useful to allow the matching process to skip a limited set of AST constructs.
 Therefore we introduce wildcards that match a set of AST constructs an arbitrary and possibly infinite number of times (zero is valid as well).
 There are two types of wildcards in our language corresponding to the two directions (horizontal and vertical) in which we can traverse the AST:
 \begin{itemize}
	\item
	The \T{wide\-Wildcard()} consumes zero to many AST elements on the same level (i.e. siblings). %
	Just like the wildcards used in regular expressions, our wildcard can be configured to match non-greedy, in which case the first successful match is returned, or greedy in which case all possible successful matches are returned.
	Skipping a limited set of AST elements can be useful since the code we are interested in could be interleaved with unrelated code e.g. logging.
	Additionally, wildcards can also be used to skip code that is not deemed to be part of the core algorithm logic.
	In the example search pattern the \T{wide\-Wildcard()} in \cref{l1:wildcard1} allows arbitrary statements at the beginning of the block.
	When matching the example prime factors implementation in \Cref{list:primefactors-impl} the used wildcards consume the definition (\Cref{lpf:listdef}) and later the write to the list holding the prime factors (\Cref{lpf:listwrite}).
	This is useful since some prime factor implementations just print the numbers directly.
	\item
	The \T{depth\-Wildcard()} matches its children anywhere within the AST starting from the current position (comparable with \T{**} in file paths).
	This is useful to specify that an AST construct can be arbitrarily nested somewhere inside an expression or statement.
	In the provided example we use this in \Cref{l1:depthWildcard} to express that the variable bound to the identifier \enquote*{index} is read somewhere on the right-hand side of the assignment since some implementations use \textit{n = n / i} instead of \textit{ n /= i}.
 \end{itemize}

\subsubsection{Alternatives}
Our DSL allows the user to specify a list of alternatives of which only one has to match by using the \T{one\-Of()} construct.
Some prime factors implementations are not interested in a complete factorization which would include the multiplicities but are content with e.g. \textit{primefactors(12) $\rightarrow$ \{2,3\}} instead of \textit{\{2,2,3\}}.
These implementations replace the while loop with an if like this: \mbox{\textit{if(n \% i == 0)\{primes.add(i); n \%= i;\}}.}
To include such implementations, we could insert a \T{one\-Of()} construct into our search pattern and specify the if as an alternative, therefore matching both possibilities.

\subsubsection{Optionals}
The \T{op\-tional()} construct present\-ed in \Cref{l1:ret} can be used to specify parts of a search pattern that may be present but are not required for a successful match.
As already mentioned, some implementations print the prime factors directly and are often methods with the return type void.
Such implementations are not required to have a return statement in Java but may include one.
Other use cases for this construct include optional optimizations in algorithm implementations and early returns.
I.e., the if statements depicted in \cref{fig:if_statements} are semantically equivalent due to the early return.
In both cases, \textit{stmt2} is only executed if the condition evaluates to false.
This essentially makes the \textit{else} optional.
We therefore provide the \T{ite().\allowbreak optional\allowbreak Otherwise(stmt)} convenience meth\-od, which matches both variants and internally uses the \T{op\-tional()} construct.

\begin{figure}
	\null\hfill\begin{minipage}[t]{1.65cm}
		\begin{lstlisting}[numbers=left,numberstyle=\tiny,basicstyle=\footnotesize{\ttfamily},numbersep=2pt,escapeinside=!!]
if(...) {     
	return;
} 
stmt2;
			\end{lstlisting}
		\end{minipage}
		\hfill
	\begin{minipage}[t]{1.65cm}
		\begin{lstlisting}[numbers=left,numberstyle=\tiny,basicstyle=\footnotesize{\ttfamily},numbersep=2pt,escapeinside=!!]
if(...) {
	return;
} else {
	stmt2;
}
		\end{lstlisting}
	\end{minipage}\hfill\null
	\caption{Two semantically equivalent if statements: one uses an early return, the other an explicit else block.}
	\label{fig:if_statements}
\end{figure}

\subsubsection{Matching Order}\label{sec:matchingOrder}
When defining a search pattern we implicitly specify the order in which we want to match the elements in the AST.
As in the example provided, this is often straightforward and simply the order in which the statements and expressions appear in the code.
However, there are cases where we want to match AST elements in \T{UN\-OR\-DERED} fashion.
One such example is method parameters which are matched \T{UN\-OR\-DERED} by default.
Another example is statements that are not dataflow dependent on each other and can therefore occur in any order inside a block.
An example of this can be seen in the code listing of \Cref{fig:exa-multiple-bindings2} in which the statements in \Cref{lmultipleBindings:stmt1} and \Cref{lmultipleBindings:stmt2} can be swapped.
To facilitate this each builder in our DSL allows the specification of the matching order for its children.
Where appropriate we expose this configuration option to the user.
One example of this is the \T{block} builder:
By using \T{block.\allowbreak has(stmt)} instead of \T{block.\allowbreak next(stmt)} all children of the block will be matched in an arbitrary order (\T{UN\-OR\-DERED}) instead of a fixed order (\T{ORDERED}).
Since the matching order is specified once for all children, the DSL prevents the user from specifying multiple matching orders meaning that either \T{block.\allowbreak has(stmt)} or \T{block.\allowbreak next(stmt)} can be used per block.
In the case of \T{UN\-OR\-DERED}, we have to try all possible combinations between the search pattern statements and AST statements during the matching.
Fortunately, many match attempts will fail immediately because the type of statements in the search pattern and the AST will often be different.

\subsection{Convenience Features}\label{sec-convenience}
\def\hl#1{\texttt{\textbf{#1}}}
	\begin{lstlisting}[float=h,label=list:pf-conv,tabsize=2,caption=A prime factors search pattern using convenience features. Changes compared to \cref{list:search-pattern-basic} in bold.,basicstyle=\scriptsize\ttfamily,language=Java,escapechar=|, numbers=left, numberstyle=\tiny, numbersep=2pt, xleftmargin=7pt]{Name}
method().bindTo("method").body(block()
|\hl{.after}|(|\hl{loop}()| |\label{l2:block:after}| |\label{l2:loop}|
		.condition(binOp().ops("<", "<="))
		.body(block().next(
			|\hl{loop}()|
				.condition(binOp().ops("%", "/")
						.lhs(varRead().bindTo("num"))
						.rhs(varRead().bindTo("index"))
				)
				.body(block()|\hl{.anywhereAfter}|(|\label{l2:block:anywhereAfter}|
						assignment()
							.lhs(varWrite().bindTo("num"))
							|\hl{.anywhereInRhs}|(varRead().bindTo("index"))|\label{l2:anywhereInRhs}|
						)))))
|\hl{.after}|(optional(returns())) |\label{l2:opt:retuns}|
	\end{lstlisting}
While the core language features presented in \cref{core-languge-primitives} are in principle sufficient to formulate even complex search patterns, our DSL benefits immensely from the convenience features that contribute to the usability of the language.
The convenience features of our DSL are methods of the builder classes which decrease the user effort for writing patterns, increase the readability and prevent errors, e.g., nesting wildcards directly.
Behind the scenes, these convenience methods use the core language features to realize the desired functionality.
\cref{list:pf-conv} shows a prime factors search pattern that makes use of the convenience methods.
In the following, we will highlight some of the convenience methods and explain their implementation:

\subsubsection{Convenience for Wildcards}
In the example in \cref{fig:primefactors-core} we had to manually specify wildcards between each pattern statement using \T{block().\allowbreak next(stmt)} to allow interleaved code.
In \cref{l2:block:after} and \cref*{l2:opt:retuns} of \cref{list:pf-conv} we now use \T{block().\allowbreak after(stmt)}, to specify that the provided pattern statements must appear \textit{somewhere} in the block, but after the last pattern statement which was added.
Since the loop is the first statement in our example it is unconstrained and may appear \textit{somewhere} in the parent block.
The optional return statement on the other hand may only appear \textit{somewhere after} the loop.
Internally, \T{block().\allowbreak after(stmt)} simply interleaves the statements with wideWildcards.

Furthermore, we use the \T{block().\allowbreak any\-where\-Af\-ter(\allowbreak stmt)} in \cref{l2:block:anywhereAfter} which is useful if we want to express that a statement may occur arbitrarily nested inside a block and be preceded by other code.
Internally, \T{block().\allowbreak any\-where\-Af\-ter(stmt)} inserts a \T{wide\-Wildcard()}, which will consume unrelated code, then wraps the contents of \T{block().\allowbreak any\-where\-Af\-ter(stmt)} in an \T{depth\-Wildcard(stmt)}.

Finally, some builders like \T{as\-sign\-ment()}, \T{var\-Def() or bin\-Op()} provide methods such as \T{anywhere\-InRhs(\allowbreak stmt)} used in \cref{l2:anywhereInRhs} to specify that the following part of the search pattern may be arbitrarily nested somewhere inside this part of the AST.
Internally, these simply wrap their contents into an \T{depth\-Wildcard(stmt)}.

\subsubsection{Convenience for Alternatives}
In order to create search patterns with a good recall it is often necessary to allow multiple equivalent alternatives for certain statements in the AST.
A good example of this is the use of the \T{loop()} builder in \cref{l2:loop}, which matches all the loop kinds that exist in Java (for, while, do-while, for-each).
Internally, \T{loop()} creates a \T{one\-Of()} of the necessary basic builders which match the individual Java loop constructs (\T{forLoop(), \T{whileLoop()}, \ldots)}.
Other examples include \T{varDefOrAss()} which matches either a variable definition or an assignment and \T{anyMod()} which matches the typical ways to modify a variable in Java, i.e., we match assignments, pre-/post-increment as well as pre-/post-decrement.

\subsubsection{Convenience for Matching Order}
As already motivated in \cref{sec:matchingOrder} it is sometimes useful to allow the matching of the AST elements in \T{UN\-OR\-DERED} fashion.
Where applicable, we expose convenience methods to enable this behavior in the DSL.
Examples include \T{method().\allowbreak hasParameters(stms\ldots)} which allows the parameters of a method to occur \T{UN\-OR\-DERED} as well as additional parameters and \T{binOp().\allowbreak compare\allowbreak Commuta\allowbreak tive()} which allows an arbitrary order of the operands which is useful for commutative operands like \T{(==, !=, *, +, \ldots)}. %

\subsection{Transforming the Search Pattern into a Pattern Tree}
\vspace{-1mm}
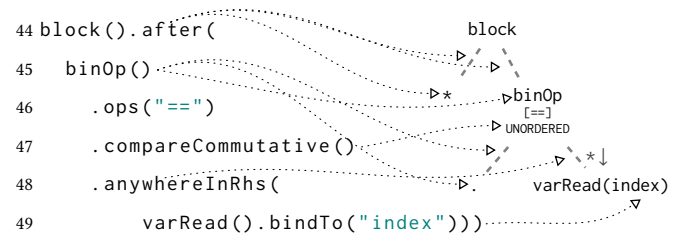
\begin{figure}[h]
	\centering
	\begin{minipage}{.55\linewidth}
  \begin{lstlisting}[basicstyle=\scriptsize\ttfamily,language=Java,escapechar=|, numbers=left, numberstyle=\tiny, numbersep=2pt,xleftmargin=7pt,lineskip=3.5pt]{Name}
|\Snode{blockcode}|block().|\Snode{after}|after(
  binOp()|\Snode{binOpcode}|
    .ops("==")|\Snode{ops}|
    .compareCommutative()|\Snode{compareCommutative}|
    |\Snode{anywhereInRhs}|.anywhereInRhs(
        varRead()|\Snode{varReadcode}|.bindTo("index"))|\Snode{bindTocode}|)
  \end{lstlisting}
  \end{minipage}\hfill
  \hspace{6mm}
\begin{minipage}{.35\linewidth}
  \scalebox{.82}{%
  \begin{forest}
    subtree,remember picture,for tree={scale=.82},
    for tree={l sep-=.55em},
    every node/.append style={transform shape}
    [\subnode{block}{block},
      [\subnode{wildcard}{*}]
      [\AsAny{\WithMod{\subnode{binOp}{binOp}}{[==]}},
        [\subnode{any}{.}]
        [\subnode{varRead}{varRead\BindTo{index}}, depthWildcardR] 
      ]
    ]
  \end{forest}}
  \begin{tikzpicture}[overlay,remember picture,every path/.append style={line cap=round}]
    \draw[-{Triangle[open,line width=.5pt]},dotted] ([xshift=4mm, yshift=0.05mm]after.north) to[out=20,in=180] ([xshift=-1.5mm, yshift=-3mm]wildcard);
    \draw[-{Triangle[open,line width=.5pt]},dotted] ([xshift=0.5mm, yshift=-0mm]binOpcode.east) to[out=-10,in=190] ([xshift=-1.3mm, yshift=-0.5mm]binOp.south west);
    \draw[-{Triangle[open,line width=.5pt]},dotted] ([xshift=0.5mm, yshift=-0mm]compareCommutative.east) to[out=10,in=180] ([xshift=-8.9mm, yshift=-7.2mm]binOp);
    \draw[-{Triangle[open,line width=.5pt]},dotted] ([xshift=1.8mm, yshift=-0mm]bindTocode.east) to[out=5,in=215] ([xshift=-2mm, yshift=0.5mm]varRead.south);

    \draw[-{Triangle[open,line width=.5pt]},dotted] ([xshift=10mm, yshift=-0.2mm]anywhereInRhs.north) to[out=10,in=196] ([xshift=-12mm, yshift=0mm]varRead.north); %
    \draw[-{Triangle[open,line width=.5pt]},dotted] ([xshift=4mm, yshift=0.05mm]after.north) to[out=20,in=180] ([xshift=2mm, yshift=2mm]wildcard);
    \draw[-{Triangle[open,line width=.5pt]},dotted] ([xshift=4mm, yshift=0.05mm]after.north) to[out=20,in=180] ([xshift=6.5mm, yshift=0.5mm]wildcard);

    \draw[-{Triangle[open,line width=.5pt]},dotted] ([xshift=0.5mm, yshift=-0mm]binOpcode.east) to[out=20,in=180] ([xshift=-3.5mm, yshift=-7.5mm]binOp.south west);
    \draw[-{Triangle[open,line width=.5pt]},dotted] ([xshift=0.5mm, yshift=-0mm]binOpcode.east) to[out=20,in=180] ([xshift=-7.1mm, yshift=-12.0mm]binOp.south west);

  \end{tikzpicture}
\end{minipage}
	\vspace{-1mm}
	\caption{Example search pattern on the left and the corresponding pattern tree on the right. The arrows indicate which part of the pattern tree is created by which part of the search pattern (obvious parts such as \T{block()} are omitted for brevity).} %
	\label{fig:search-pattern-to-pattern-tree}
\end{figure} %
Before starting the matching process, we first transform the search pattern, specified by the user in our embedded Java DSL, into a different representation called the pattern tree.
This representation facilitates the matching process since we can now \enquote{simply} compare the structure and attributes of the pattern tree with those of the AST.
Additionally, this facilitates the separation of concerns.
While the builders are responsible for specifying and building the search pattern, the pattern tree is concerned with the matching of the pattern.
The pattern tree is created by calling the \T{build()} method of a builder object.
As is typical of the builder design pattern, each builder knows how to translate his configuration into a pattern tree.
To obtain the complete pattern tree the transformation is also delegated to all the children of a builder.
During the transformation, we translate the convenience features described in \cref{sec-convenience} into the core features of our language, register specified bindings,
determine the matching order for the children and generate predicates to check attributes of the AST such as the operator type.
\Cref{fig:search-pattern-to-pattern-tree} illustrates an example of the transformation.
The builders themselves generate a pattern tree node that matches exactly the Java construct described.
Convenience methods like \T{af\-ter(stmt)} and \T{anywhere\-InRhs(stmt)} additionally generate the necessary wildcards.
In the figures these are represented as \enquote{{\ttfamily *}} and \enquote{{\footnotesize\ttfamily*\(\downarrow\)}} for the \T{wide\-Wildcard()} and \T{depth\-Wildcard()} respectively.
The \T{ops(stmt\ldots)} generates the necessary predicate to check the operator kind while the \T{bind\-To()} method registers the binding constraints.
Because of the \T{compare\-Commutative()} specification, we set the matching order of the \T{bin\-Op()} children to \T{UN\-OR\-DERED}.
Finally, since no left-hand side was specified, we generate an \T{any()} represented by the~\enquote{{\ttfamily .}} to accept any expression on the left-hand side.

\subsection{Matching}
\subsubsection{Matching State}
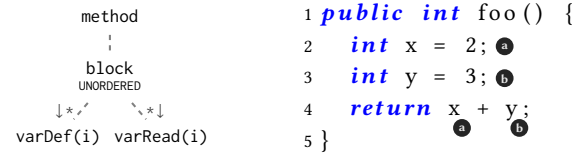
\begin{figure}[h]
	\centering\begin{minipage}{.45\linewidth}
	\centering
	\scalebox{.65}{
	\begin{forest}
		subtree tight,
		for tree={l sep-=.55em}
		[method
			[\AsAny{block}
				[varDef\BindTo{i},depthWildcardL]
				[varRead\BindTo{i},depthWildcardR]
			]
		]
	\end{forest}}
	\end{minipage}\hfill\begin{minipage}{.45\linewidth}\small
	\begin{lstlisting}[numbers=left,numberstyle=\tiny,basicstyle=\footnotesize{\ttfamily},numbersep=2pt]
public int foo() {
	int x = 2;!*\Snode{variantA}*!!*\label{lmultipleBindings:stmt1}*!
	int y = 3;!*\Snode{variantB}*!!*\label{lmultipleBindings:stmt2}*!
	return x!*\Snode{variantA2}*! + y!*\Snode{variantB2}*!;
}
	\end{lstlisting}
	\end{minipage}%
	\begin{tikzpicture}[overlay,remember picture,matchextraopts/.style={scale=.66}]
		\Match([xshift=1mm,yshift=2.36mm]variantA.south east){a};
		\Match([xshift=-.85mm]variantA2.south east){a};
		\Match([xshift=1mm,yshift=2.36mm]variantB.south east){b};
		\Match([xshift=-.85mm]variantB2.south east){b};
	\end{tikzpicture}
	\caption[Example showcasing two possible successful matching states.]{Example showcasing two possible successful matching states. Either \begin{orlist}
		\item we bind \T{\textit{i}} to the variable \T{x}
		\item the variable \T{y}.
	\end{orlist}}
	\label{fig:exa-multiple-bindings2}
	\end{figure}

During the matching, we store the current matching state in an immutable data structure and record:
\begin{itemize}
	\item \textit{metadata}, like the current depth of the matching
	\item \textit{nodes already matched} as a bidirectional mapping between the nodes of the pattern tree and the AST
	\item \textit{bindings} which represent the mapping between the identifiers created by the binding constraints and the AST elements bound to them.
\end{itemize}
Each node of our search pattern has access to the current matching state (e.g., to check the binding constraints) and returns a (potentially empty) set of new matching states (e.g., one for each possible binding). %
Allowing to return more than one matching state is important, as illustrated by \cref{fig:exa-multiple-bindings2}.
With the pattern tree on the left, we allow the statements of the \T{block()} to match in \T{UN\-OR\-DERED} fashion.
This leads to two valid matching states for \T{varDef().bindTo("i")}---which searches for the definition of a variable---binding the declarations in either line~2 or line~3.
This in turn influences what can be matched by \T{var\-Read().bindTo("i")} in combination with the \T{depth\-Wildcard()}, either the left-hand or the right-hand side of \T{x + y}.
We consider a matching state to be valid if no match failed and if all binding constraints map to a non-empty set of matched parts.
\subsubsection{Matching Procedure}
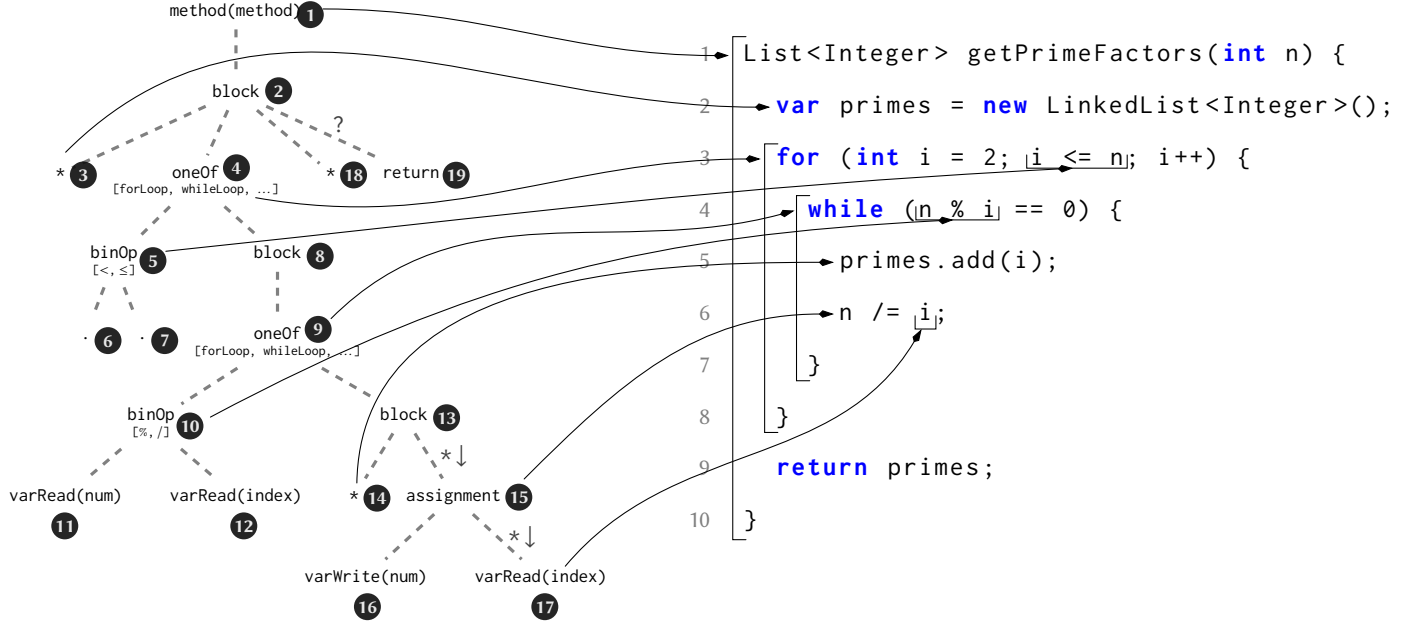
\begin{figure*}
	\centering
	\begin{minipage}{.45\linewidth}
\begin{forest}
  subtree,remember picture,for tree={scale=.65,s sep-=.5em,l sep-=.55em},
  every node/.append style={transform shape}
  [\subnode{defBody}{method\BindTo{method}},
    [\subnode{block2p}{block},
      [\subnode{w3}{*}]
      [\WithMod{\subnode{oneOf4}{oneOf}}{[forLoop, whileLoop, \ldots]},
        [\WithMod{\subnode{binOp5}{binOp}}{[\(<\),\,\(\leq\)]},
          [\subnode{dot1}{.}]
          [\subnode{dot2}{.}]
        ]
        [\subnode{block8}{block},
          [\WithMod{\subnode{oneOf9}{oneOf}}{[forLoop, whileLoop, \ldots]},
              [\WithMod{\subnode{binOp10}{binOp}}{[\(\%\),\,\(/\)]},
                [\subnode{varRead11}{varRead\BindTo{num}}]
                [\subnode{varRead12}{varRead\BindTo{index}}]
              ]
              [\subnode{block13}{block},
                [\subnode{w14}{*}]
                [\subnode{ass15}{assignment}, depthWildcardR
                  [\subnode{varWrite16}{varWrite\BindTo{num}}]
                  [\subnode{varRead17}{varRead\BindTo{index}}, depthWildcardR]
                ]
              ]
          ]
        ]
      ]
      [\subnode{w18}{*}]
      [\subnode{return19}{return}, optional]
    ]
  ]
\end{forest}
\end{minipage}\hfill
\begin{minipage}{.45\linewidth}
{\small\begin{lstlisting}[lineskip=6.5pt]
!*\Snode{method}*!List<Integer> getPrimeFactors!*\Snode{PrimeFactors}*!(int n) !*\Snode{block}*!{
  !*\Snode{block-inner}*!var primes = new LinkedList<Integer>();
  !*\Snode{for}*!for (int i = 2; !*\Snode{cmp}*!i <= n!*\Snode{cmp@}*!; i++) !*\Snode{block2}*!{
    !*\Snode{block-inner2}*!!*\Snode{while}*!while (!*\Snode{mod}*!n % i!*\Snode{mod@}*! == 0) !*\Snode{block3}*!{
      !*\Snode{block-inner3}*!primes.add(i);
      !*\Snode{anyMod}*!!*\Snode{anyVar}*!n!*\Snode{anyVar@}*! /= !*\Snode{varReadi}*!i!*\Snode{varReadi@}*!!*\Snode{anyMod@}*!;
    !*\Snode{block3@}*!!*\Snode{while@}*!}
  !*\Snode{block2@}*!!*\Snode{for@}*!}
  return primes;
!*\Snode{block@}*!!*\Snode{method@}*!}
\end{lstlisting}}
\begin{tikzpicture}[overlay,remember picture,every path/.append style={line cap=round}]
  \draw ([xshift=.5mm,yshift=1mm]method.north) -- ++(-1.75mm,0) coordinate[yshift=-2.5mm] (@target) |- ([xshift=.5mm,yshift=-1mm]method@.south);
  \draw[-Kite] ([xshift=-8mm]defBody.east) to[out=0,in=180] ([xshift=-.5mm]@target);
  \draw ([xshift=.5mm,yshift=.5mm]for.north) -- ++(-1.75mm,0) coordinate[yshift=-2mm] (@target) |- ([xshift=.5mm,yshift=-.5mm]for@.south);
  \draw[-Kite] ([xshift=2mm,yshift=-1.25mm]oneOf4.south)  to[out=350,in=180] ([xshift=-.5mm]@target);
  \path (binOp5.south) -- (cmp@.south) coordinate[pos=.5] (@);
  \draw ([yshift=2mm,xshift=-.25mm]cmp.south west) -- ++(0,-1.75mm) -| ([xshift=.25mm,yshift=2mm]cmp@.south east) coordinate[pos=.25] (@target);
  \draw[-Kite] ([xshift=-1mm,yshift=0mm]binOp5.east) to[out=6,in=183] ([xshift=-.5mm]@target);
  \draw[-Kite] ([xshift=-2mm,yshift=0mm]w3.north east) to[out=45,in=180] ([xshift=-.5mm]block-inner);
  \draw ([xshift=.5mm,yshift=.5mm]while.north) -- ++(-1.75mm,0) coordinate[yshift=-2mm] (@target) |- ([xshift=.5mm,yshift=-.5mm]while@.south);
  \draw[-Kite] ([xshift=-5mm,yshift=-1.25mm]oneOf9.north east)  to[out=45,in=195] ([xshift=-1.5mm]while);
  \path (binOp10.east) -- (mod@.south) coordinate[pos=.5] (@);
  \draw ([yshift=2mm,xshift=-.15mm]mod.south west) -- ++(0,-1.75mm) -| ([xshift=.15mm,yshift=2mm]mod@.south east) coordinate[pos=.25] (@target);
  \draw[-Kite] ([xshift=0mm,yshift=-3.5mm]binOp10.north east) to[out=28,in=183] ([xshift=-.5mm]@target);
  \draw[-Kite] ([xshift=0mm,yshift=-1mm]w14.north) to[out=89,in=180] ([xshift=-.5mm]block-inner3);
  \draw[-Kite] ([xshift=-5mm,yshift=-1mm]ass15.north east) to[out=45,in=180] ([xshift=-.5mm]anyMod);

  \path (varRead17.north) -- (varReadi@.south) coordinate[pos=.5] (@);
  \draw ([yshift=1.25mm,xshift=-.25mm]varReadi.south west) -- ++(0,-1.75mm) -| ([xshift=.25mm,yshift=1.25mm]varReadi@.south east) coordinate[pos=.25] (@target);
  \draw[-Kite] ([xshift=-1.5mm,yshift=-1.85mm]varRead17.north) to[out=50,in=239] ([xshift=-.5mm]@target);

  \Match([xshift=-1.1cm,yshift=3.5mm]defBody.south east){1}; 
  \Match([yshift=3.5mm,xshift=-3.5mm]block2p.south east){2};  
  \Match([yshift=3mm,xshift=-0.95mm]w3.south east){3};
  \Match([yshift=4mm,xshift=-8mm]oneOf4.south east){4};
  \Match([yshift=3mm,xshift=-4mm]binOp5.south east){5};
  \Match([yshift=2.5mm,xshift=-1mm]dot1.south east){6};
  \Match([yshift=2.5mm,xshift=-1mm]dot2.south east){7};
  \Match([yshift=3mm,xshift=-3.5mm]block8.south east){8};
  \Match([yshift=4mm,xshift=-8mm]oneOf9.south east){9};
  \Match([yshift=2.5mm,xshift=-4mm]binOp10.south east){10};
  \Match([yshift=0mm,xshift=-5.5mm]varRead11.south){11};
  \Match([yshift=0mm,xshift=-5mm]varRead12.south){12};
  \Match([yshift=3mm,xshift=-3.5mm]block13.south east){13};
  \Match([yshift=3mm,xshift=-0.95mm]w14.south east){14};
  \Match([yshift=3.8mm,xshift=-7mm]ass15.south east){15};
  \Match([yshift=0mm,xshift=-5.5mm]varWrite16.south){16};
  \Match([yshift=0mm,xshift=-5.5mm]varRead17.south){17};
  \Match([yshift=3mm,xshift=-0.95mm]w18.south east){18};
  \Match([yshift=3mm,xshift=-4.5mm]return19.south east){19};
\end{tikzpicture}
\end{minipage}\\[.5ex]
	\vspace{0.5em}
	\caption{Example pattern tree for the prime factors search pattern from \Cref{list:pf-conv} on the left and an example implementation which would satisfy the pattern tree on the right. The numbers specify the matching order and the arrows of selected pattern tree nodes indicate which code fragment is matched by them.}
	\label{fig:matching-graphic-new}
\end{figure*}
The matching process starts with a search pattern and a set consisting of a single initial matching state.
All nodes of the source code AST that match the root node type of our search pattern are valid entry points for the search.
For each match between a pattern node and an AST node, we update the set with the newly produced matching states.
We then continue by trying to match each child of the pattern node with the next AST node and every valid matching state in the state set.
If there are no nodes left to match in the pattern specification (some nodes like \T{one\-Of()} can match even if some of their children fail), we consider each valid matching state a successful match.

To illustrate the matching process, we reconstruct the successful matching shown in \cref{fig:matching-graphic-new}.
We start the search with the \T{method()} (root-)node considering only methods and constructors as starting points.
For this example, we only match the \T{getPrimeFactors(int)} method and bind the matched node to the \T{\textit{method}} identifier (producing one valid matching state).
We continue by matching the child node \T{block()} with the syntactical block element (again producing one valid matching state).
The matching continues with the \T{wide\-Wildcard()} in step 3 which creates four possible matching states by consuming either
\begin{orlist}
	\item nothing
	\item only the statement in line 2
	\item the statements in lines 2-8
	\item all the statements in the block (lines 2-9).
\end{orlist}
Of these possible matching states only the second one will lead to a successful match, since in the other cases the rest of the search pattern can not be satisfied.
In step 4 the \T{one\-Of()} construct tries all of its possible alternatives, of which only the \T{for\-Loop()} matches.
We carry on by matching the \T{bin\-Op()} in step 5 which satisfies the required operator kind.
Both of the \T{any()} children match (against \T{i} and \T{n} respectively), which means that we continue the matching with the \T{block()} in step 8.
In step 9 the second \T{one\-Of()} also tries all of the loop kinds in this instance succeeding with \T{whileLoop()}.
In step 10 we check the condition which satisfies the required operator kind and match both \T{varRead()} children against \T{i} and \T{n} respectively, binding the variables in the process.
The matching continues with the loop body matching the \T{block()} in step 13.
The contained \T{wide\-Wildcard()} creates three matching states in step 14 by consuming either
\begin{orlist}
	\item nothing
	\item only the statement in line 5
	\item all the statements in the block (lines 5-6)
\end{orlist}
of which again only the second one will result in a successful match.
In step 15 we proceed with matching the \T{as\-sign\-ment()} and its children which both check and satisfy the binding constraints.
We then return to matching the method body with steps 18 and 19.
Here the combination of a \T{wide\-Wildcard()} and the \T{op\-tional()} return statement, represented by the \enquote{{\ttfamily ?}} in the pattern tree, leads to multiple matching possibilities:
\begin{itemize}
	\item The \T{wide\-Wildcard()} can consume nothing, with the optional \T{re\-turns()} successfully matching the last statement.
	\item The \T{wide\-Wildcard()} can consume the return statement, while the optional \T{re\-turns()} matches nothing. This is also a valid match since the \T{re\-turns()} is optional.
	\item Both the \T{wide\-Wildcard()} and the optional \T{re\-turns()} can refrain from matching, since both are not required to consume anything for a successful match.
\end{itemize}
All three successful matching states cover the range of the method (lines 1-10) which is matched by the \T{method()} root-node.
We therefore simply unify all three matching states because the user is most likely only interested in the fact that we found an implementation of the prime factors algorithm.

\section{Evaluation}\label{sec-evaluation}%
	We are interested in exploring how accurately algorithms can be recognized based solely on the program AST.
	We therefore evaluate the accuracy of our prototype called AlDeSCo (Algorithm Detection on Source Code) and formulate the following research questions:
	\begin{enumerate}[label={\textbf{RQ\arabic*:}},ref={RQ\arabic*},leftmargin=*]
		\item \label{rq:1}How well can algorithms be detected using our AST-based pattern language? %
		\item \label{rq:2}How do our results compare to large language models?
		\item \label{rq:3}How do our results compare to code clone detection tools?
	\end{enumerate}

\subsection{Method}\label{section:method}  %
		We first evaluate the recognition performance of our search patterns on the BigCloneEval~\cite{svajlenko2016bigcloneeval} benchmark since it provides several labeled and manually verified algorithm implementations from real-world source code bases.
		Then we compare our results with Codellama, a large language model that has successfully been used on other code-related tasks~\cite{roziere2024codellama, liu2024LLMCodeGenEval, silva2023repairllama}.
		Afterward, we compare our results with multiple CCD tools as a baseline.
		We chose this comparison since different implementations of an algorithm can be considered clones of each other.
		Consequently, a CCD tool provided with a reference implementation of an algorithm can be used to find clones in a code base i.e. other implementations of the same algorithm.
		Depending on the similarity of the code clones they can be separated into the following types:
		\begin{description}[labelindent=\parindent, leftmargin=1.125cm] %
			\item[T1:] Differ only in whitespace, layout, and comments.
			\item[T2:] Differ additionally in identifiers, literals, and types.
			\item[T3:] Differ additionally in statements (e.g., added,~\ldots).
			\item[T4:] Differ in syntax but perform the same computation.
		\end{description}
		The BigCloneEval~\cite{svajlenko2016bigcloneeval} dataset is quite diverse with many Type-3 and Type-4 clones, strengthening our evaluation.

The IJaDataset~2.0 contains over 365\,\textsc{mloc} of Java code and is the basis for both the BigClone\-Bench~\cite{Svajlenko2014} and BigCloneEval~\cite{svajlenko2016bigcloneeval} benchmarks.
		To create BigClone\-Bench Svajlenko et al. defined~43 different functionalities such as \textit{Web Download}, \textit{Send E-Mail} or \textit{Copy File}.
		They then created a regular expression (regex) for each functionality to mine the IJaDataset for candidate methods implementing said functionality.
		These candidates were then manually verified by a set of judges resulting in true positive methods implementing the functionality and false positive methods which don't.
		The true positives were then paired up to create clone pairs~\cite{Svajlenko2014}.

BigCloneBench~\cite{Svajlenko2014} comprises the IJaDataset, additional example implementations for the functionalities, the labels of the validated methods and the clone pair information.

Later, Svajlenko et al. created BigCloneEval~\cite{svajlenko2016bigcloneeval}, that builds on BigCloneBench~\cite{Svajlenko2014} to evaluate CCD tools automatically.
		BigCloneEval reduces the storage and runtime requirements by only including the subset of files that contain labeled methods.
		Nonetheless running a tool on BigCloneEval is equivalent to running it on BigCloneBench in terms of measuring recall~\cite{svajlenko2016bigcloneeval}.
		This is the case since only the files with the labeled methods are required to calculate the recall.
		BigCloneEval is used by many CCD tools~\cite{Saini2018} making it a popular benchmark for their comparison.

Not all functionalities contained in BigCloneEval would generally be considered to be an algorithm.
		Therefore we selected a subset consisting of \textit{Prime Factors}, \textit{Greatest Common Divisor (GCD)}, \textit{Fibonacci}, \textit{Palindrome}, \textit{Bubble Sort} and \textit{Binary Search} for our experiments.

We then created a search pattern for each of the selected algorithms using our DSL.
		First, we performed a web search to retrieve a reference implementation of the algorithm in established sources including coding and Q\&A sites.
		Then we used our pattern language to express the key features of the reference implementation in our search pattern and benchmarked its performance on BigCloneEval.
		Depending on the algorithm we would repeat this process and use up to three reference implementations to derive our search pattern.
		To avoid bias in the results we did not use any implementations from the dataset to create our search patterns and also refrained from analyzing any methods in the dataset until after the final evaluation.
		A replication package for the results of our prototype is available on Zenodo~\cite{neumueller2024Zenodo}.

\subsection{Evaluation on BigCloneEval}\label{sec:EvalBCE} %

\begin{table}
	\centering
	\def\mtfrmt#1{(\(\num[round-mode=none]{#1}\) methods total)}
\def\AlgoBanner#1{\multicolumn{10}{l}{{\large\strut}\rlap{\strut#1}}}%
\def\TableHeader{TP&FP&FN%
&MCC&{\FMeasure}&TPR&PPV& [h:mm:ss] & \#MS}
\newcolumntype{K}{S[table-format=1.2,round-mode=places]} %
\newcolumntype{M}{S[round-mode=places,round-precision=1,exponent-mode = scientific,table-number-alignment=right,table-alignment-mode=none,output-exponent-marker={e}]}
\makeatletter
\def\TimeFormat#1:#2:#3\;{%
   {#1}:%
   \ifdim#2pt<10pt 0\fi\num[round-precision=0,round-mode=places]{#2}:%
   \ifdim#3pt<10pt 0\fi\num[round-precision=0,round-mode=places]{#3}%
}
\def\TableLine "fMeasure": #1,
"recall": #2,
"precision": #3,
"accuracy": #4,
"mcc": #5,
"truePositives": #6,
"falsePositives": #7,
"falseNegatives": #8\; #9\s{%
  We & #6 & #7 & #8 %
  & #5 & #1 & #2 & \ifnum#6=0 {\tiny N/A~}\else #3\fi & \ifstrequal{#9}{?}{??}{\expandafter\TimeFormat#9\;}\ignorespaces
}
\def\SameAsAbove{\color{gray}}
\robustify\SameAsAbove
\def\TableLineB "fMeasure": #1,
"recall": #2,
"precision": #3,
"accuracy": #4,
"mcc": #5,
"truePositives": #6,
"falsePositives": #7,
"falseNegatives": #8\; #9\s{%
  #6 & #7 & \SameAsAbove#8 %
  & #5 & #1 & #2 & \ifnum#6=0 {\tiny N/A~}\else #3\fi & \SameAsAbove\expandafter\TimeFormat#9\;\ignorespaces
}

\def\TableLineC "fMeasure": #1,
"recall": #2,
"precision": #3,
"accuracy": #4,
"mcc": #5,
"truePositives": #6,
"falsePositives": #7,
"falseNegatives": #8\; #9\s{%
\SameAsAbove#6 & \SameAsAbove#7 & \SameAsAbove#8 %
  & \SameAsAbove#5 & \SameAsAbove#1 & \SameAsAbove#2 & \ifnum#6=0 {\tiny N/A~}\else \SameAsAbove#3\fi & \SameAsAbove\expandafter\TimeFormat#9\;\ignorespaces
}
\def\TablePrimeFactors{\AlgoBanner{Prime Factors \mtfrmt{1302}}\\ %
\TableLine "fMeasure": 0.8181818181818182,
"recall": 0.8571428571428571,
"precision": 0.782608695652174,
"accuracy": 0.9938556067588326,
"mcc": 0.8159359231,
"truePositives": 18,
"falsePositives": 5,
"falseNegatives": 3\; 0:0:04.15\s & 13676\\
}

\def\TableFibonacci{%
   \AlgoBanner{Fibonacci \mtfrmt{3608}}\\
\TableLine "fMeasure": 1.0,
"recall": 1.0,
"precision": 1.0,
"accuracy": 1.0,
"mcc": 1.0,
"truePositives": 212,
"falsePositives": 0,
"falseNegatives": 0\; 0:0:04.967\s & 140231\\
}

\def\TableBinarySearch{
   \AlgoBanner{Binary Search \mtfrmt{57890}}\\
\TableLine "fMeasure": 0.34137291280148424,
"recall": 0.228287841191067,
"precision": 0.6764705882352942,
"accuracy": 0.9938676800829159,
"mcc": 0.3907519143,
"truePositives": 92,
"falsePositives": 44,
"falseNegatives": 311\; 0:10:33.630\s & 19538533\\
}

\def\TableBubbleSort{%
\AlgoBanner{Bubble Sort \mtfrmt{23454}}\\
\TableLine "fMeasure": 0.8733153638814016,
"recall": 0.9700598802395209,
"precision": 0.7941176470588235,
"accuracy": 0.9979960774281572,
"mcc": 0.8767560588,
"truePositives": 162,
"falsePositives": 42,
"falseNegatives": 5\; 0:0:37.245\s & 382747\\
}

\def\TableGreatestCommonDivisor {
\AlgoBanner{GCD \mtfrmt{2024}}\\
\TableLine "fMeasure": 0.4444444444444444,
"recall": 0.6086956521739131,
"precision": 0.35,
"accuracy": 0.982707509881423,
"mcc": 0.4536446563,
"truePositives": 14,
"falsePositives": 26,
"falseNegatives": 9\; 0:0:40.96\s & 408352\\
}

\def\TablePalindrome{%
\AlgoBanner{Palindrome \mtfrmt{3735}}\\
\TableLine "fMeasure": 0.9634146341463415,
"recall": 0.9404761904761905,
"precision": 0.9875,
"accuracy": 0.9967871485943776,
"mcc": 0.9620485525,
"truePositives": 158,
"falsePositives": 2,
"falseNegatives": 10\; 0:0:5.827\s & 72988\\
}

	\def\As{\(\widehat{=}\)}
	\caption{Results of the Benchmark with BigCloneEval and comparison with Codellama}
	\label{result-table}
	\resizebox\linewidth!{\begin{tabular}{>{\color{gray}}lr@{\hspace{6pt}}r@{\hspace{6pt}}r@{\hspace{6pt}}K@{\hspace{6pt}}K@{\hspace{6pt}}K@{\hspace{6pt}}K@{\hspace{6.5pt}}r@{\hspace{6pt}}M}
	\hline
	\vphantom{\(I_I^I\)} & \TableHeader \\
	\hline
	\hline
	\TablePrimeFactors
	Llama & 20 & 190 & 1 & 0.275392013051341 & 0.173160173160173 & 0.952380952380952 & 0.0952380952380952 & 0:06:24 & {\tiny N/A~} \\ %
	\TableGreatestCommonDivisor
	Llama & 20 & 60 & 3 & 0.456726837400324 & 0.388349514563106 & 0.869565217391304 & 0.25 & 0:11:50 & {\tiny N/A~} \\
	\TableFibonacci
	Llama & 209 & 80 & 3 & 0.833692832281298 & 0.834331337325349 & 0.985849056603773 & 0.72318339100346 & 0:24:13 & {\tiny N/A~} \\
	\TablePalindrome
	Llama & 166 & 334 & 2 & 0.544433138495271 & 0.497005988023952 & 0.988095238095238 & 0.332 & 0:37:34 & {\tiny N/A~} \\
	\TableBubbleSort
	Llama & 162 & 2233 & 5 & 0.242738693727899 & 0.126463700234192 & 0.97005988023952 & 0.0676409185803757 & 1:58:10 & {\tiny N/A~} \\
	\TableBinarySearch
	Llama & 396 & 9226 & 7 & 0.183620749042183 & 0.079002493765586 & 0.982630272952853 & 0.0411556848887965 & 4:18:50 & {\tiny N/A~} \\
	\hline
	\end{tabular}}
	\begin{center}
		\scriptsize TP=True positives, FP=False positives, FN=False negatives, MCC=Matthews correlation coefficient, F\textsubscript{1}=\FMeasure-Score, TPR=Recall, PPV=Precision, [h:mm:ss]=Time in hours, \#MS=Total \# of matching states
	\end{center}
	\end{table}

Usually, BigCloneBench and by extension, BigCloneEval are not used to evaluate precision~\cite{Farmahinifarahani2019, svajlenko2016bigcloneeval, SvajlenkoSurveyCCDBenchmarking}.
	Because the IJaDataset was mined by using a regex search as a heuristic it is likely that many methods implementing the functionality were not found and therefore not labeled.
	This means that some of the \enquote*{false positives} reported by a tool could be correct implementations with a missing label.
	It is however possible to calculate a lower bound for the precision by assuming that each detected method unknown to the benchmark is a false positive~\cite{Svajlenko2014}.
	We have adopted this strategy to evaluate our search patterns with BigCloneEval in the following.

\cref{result-table} contains the results of our experiments.
	We calculated Matthews correlation coefficient~\cite{chicco2020advantages}, \FMeasure score, precision, and recall.
	Since the precision we calculate is a lower bound, the \FMeasure score is a lower bound as well.
	Additionally, we report the runtime for each search pattern, the number of matching states produced and the number of methods the functionality consists of.
	Note that \enquote{methods} in this context refers to all Java executables, including methods, constructors and anonymous blocks.

Our search patterns for \textit{Prime Factors}, \textit{Fibonacci}, \textit{Palindrome} and \textit{Bubble Sort} achieve good results both in terms of precision and recall on the dataset.
	Achieving this result was easy in the case of \textit{Fibonacci}.
	The reason behind this is BigCloneEval which only contains very similar implementations of a recursive \textit{Fibonacci} implementation.
	Since matching the recursive version already produced near-perfect results we did not extend the \textit{Fibonacci} search pattern to match iterative implementations as well.
	This could result in a vastly different performance on other datasets.
	However, as we will discuss in more detail later, the implementations of the other algorithms in the dataset are more diverse and contain many Type-3 and Type-4 clones making their detection more challenging.
	We therefore had to analyze multiple example implementations in order to arrive at a good search pattern.
	While many implementation differences can be abstracted away using the convenience features of our pattern language sometimes this is not desirable.
	Typically our search patterns use the \T{one\-Of()} construct to incorporate an iterative and recursive version of the algorithm.
	Splitting the search pattern into an iterative and recursive part is useful since it is often desirable to require and use the (additional) parameters that are only present in recursive implementations.

The \textit{GCD} search pattern achieves mixed results.
	While this search pattern has an acceptable recall it is too permissive resulting in a non-ideal precision.

For \textit{Binary Search} we manage to achieve acceptable precision but fail to detect a considerable amount of implementations.
	We believe that our reliance on just two reference implementations for this pattern resulted in the oversight of a common implementation variant, which could explain this result.
	Additionally, while all other search patterns finished in under a minute, this search pattern took approximately 11 minutes.
	The \textit{Binary Search} functionality contains the most algorithm candidates with \num{57890} methods.
	However, it turned out that the significant increase in runtime was caused by a very small number of methods.
	These methods are very large spanning hundreds of lines.
	They contain the abstractly defined structure of the search pattern.
	Additionally, they contain a huge number of variable declarations.
	Due to the way we formulated our search pattern the matching algorithm has to try out all possible combinations of variable bindings in order to match the algorithm.
	Therefore, the number of generated matching states significantly increases leading to a long runtime.
	This is a general problem with our approach and the AST representation.
	Methods to tackle this problem are discussed in~\cref{section:future-work}.

\AnswerBox{1}{
		The results indicate that our AST-based approach is suited for algorithm recognition.
		We were able to create search patterns with solid performance for most algorithms.
		However, \textit{Binary Search} shows that further improvements to our approach are necessary.
	}

\subsection{Comparison with Large Language Models}\label{sec:Comparison-LLMs}
Large language models are massive machine learning models consisting of billions of parameters.
By training them on extensive corpora of text these models acquire impressive language processing skills enabling them to solve a variety of tasks~\cite{zhao2023surveyLLMs}.

In the recent past, LLMs have successfully been used on a variety of code-related tasks such as code summarization~\cite{2022LLMsCodeSummarization, 2023transformersCodeSummarization}, bug fixing~\cite{2023APR_LLMs, 2023RepairIsNearlyGeneration}, code generation~\cite{2021EvalLLMCodeGen, liu2024LLMCodeGenEval} and code clone detection~\cite{dou2023understanding, 2024LLM_CCD}.
This motivates us to investigate the application of LLMs to algorithm recognition, which to the best of our knowledge has not been explored in previous literature.

We choose Codellama-Instruct~\cite{roziere2024codellama} for our evaluation. Codellama is based on the open-source Llama2 model architecture which has shown great performance in many tasks~\cite{taori2023alpaca, zhao2023surveyLLMs}.
Codellama-Instruct has been finetuned for code understanding as well as following instructions~\cite{roziere2024codellama}, making it an ideal candidate for the task at hand.
We use the seven billion parameter model which was quantized from 16 to 8 bit, in order to reduce model size with a minimal reduction in quality loss~\cite{zhao2023surveyLLMs}.
All experiments were run on a high-performance cluster node utilizing a NVIDIA H100 GPU.

To evaluate the LLM on BigCloneEval, we iterated over every method in the dataset and prompted the LLM if the method was an implementation of the algorithm.
From our preliminary experiments, we concluded that Codellama would not perform well using a simple prompt like \texttt{"Is the snippet an implementa\-tion of the algorithm?"}.
We, therefore, applied in-context learning~\cite{brown2020ICL} and provided a positive and negative example with the correct responses.
This significantly improved the results of Codellama.
For the positive example, we used one of the reference implementations from the internet, which were also used to derive our search patterns. 
The negative example was a simple method for choosing a password consisting of \T{n} random letters.
Our complete prompt would therefore look like this:
{\small %
\begin{lstlisting}[breaklines=true,breakatwhitespace=true,stringstyle={\textit},showstringspaces=false,numbers=none,breakindent=0pt,escapeinside=!!]
!\textbf{\mbox{System Message}}!: "You are a helpful, respectful and honest assistant." 
!\textbf{\mbox{(Provided) User Prompt}}!: "SNIPPET: !\color{gray}\{positive\_example\}! Does the snippet implement !\color{gray}\{algorithm\}!, only answer with 'Yes' or 'No'?"
!\textbf{\mbox{(Provided) LLM Answer}}!: "Yes"
!\textbf{\mbox{(Provided) User Prompt}}!: "SNIPPET: !\color{gray}\{negative\_example\}! Does the snippet implement !\color{gray}\{algorithm\}!, only answer with 'Yes' or 'No'?"
!\textbf{\mbox{(Provided) LLM Answer}}!: "No"
!\textbf{\mbox{(Evaluation) User Prompt}}!: "SNIPPET: !\color{gray}\{method\}! Does the snippet implement !\color{gray}\{algorithm\}!, only answer with 'Yes' or 'No'?"
\end{lstlisting}
}
Based on the algorithm, we populated the placeholders with appropriate values e.g. replacing \textcolor{gray}{\{algorithm\}} with \textit{\enquote{Fibonacci}}.
We are interested in obtaining factually correct answers rather than diverse LLM responses.
Therefore we follow Dou et al.~\cite{dou2023understanding} and set the temperature~\cite{ackley1985learning}, Top-p (i.e., Nucleus Sampling~\cite{holzman2019NucleusSampling}), and Top-k~\cite{Fan2018Topk} parameters to 0.2, 0.1 and 10 for inference.
Given the probabilistic nature of outputs generated by LLMs, we report the results of Codellama as an average over three separate runs.

Table \ref{result-table} presents a comparison between the results of Codellama and our search patterns.
In general, Codellama is able to find almost all of the true positives leading to a very good recall across all algorithms.
Unfortunately, it also produces many false positives which heavily reduces its overall \FMeasure-score.
With the exception of \textit{Fibonacci} the number of false positives always exceeds the number of true positives.
As mentioned earlier, BigCloneEval contains only very similar implementations of a recursive \textit{Fibonacci} implementation.
We provided Codellama with exactly such an implementation as the positive example for \textit{Fibonacci} which could explain this result.
Codellama also achieves respectable \FMeasure-scores for \textit{GCD} and \textit{Palindrome} but the results for \textit{Prime Factors}, \textit{Bubbe Sort} and \textit{Binary Search} are subpar.
Our search patterns achieve a better \FMeasure-score while also being faster to execute.
\AnswerBox{2}{
	The results show that our search patterns are significantly more precise than the LLM while achieving similar recall, resulting in a better \FMeasure-score than the LLM.
	Additionally, the LLM has a significantly longer runtime, requiring between 17 to 376 times longer than our approach.
}

\subsection{Comparison with CCD Tools}\label{sec:Comparison-CCD-Tools}
	\begin{table}
		\centering
		\def\mtfrmt#1{}
\def\NaN{{\color{lightgray}\tiny N/A~}}
\def\AlgoBanner#1{%
   \vphantom{\large\(I^1\)}\multirow{6}{=}{\rotatebox[origin=c]{90}{#1\hspace*{-4pt}}}%
}%
    \sisetup{table-format=1.2,round-mode=places}%
\def\TableLine "tp": #1,
    "fn": #2,
    "total": #3,
    "recall": #4\;{%
   \strut\ifnum#1=0\ifnum#2=0\color{lightgray}\fi\fi\num[round-mode=none]{#3} & %
    \ifnum#1=0 %
      \ifnum#2=0 \NaN\else #4\fi
   \else #4\fi
}

\def\Same{\color{gray}}

\def\TableHeader{{\large\strut} & TP & FN & Total & TPR}
\def\TableFibonacci{%
\AlgoBanner{Fibonacci \mtfrmt{3608}}
& T1 & \TableLine 
"tp": 0,
"fn": 0,
"total": 0,
"recall": null\; & \NaN & \NaN & \NaN & \NaN & 11194 & 1 \\
& T2 & \TableLine 
"tp": 1,
"fn": 0,
"total": 1,
"recall": 1\; & 1.0 & 1.0 & 0 & 1 & \Same1 & \Same1 \\
& VST3 & \TableLine 
"tp": 0,
"fn": 0,
"total": 0,
"recall": null\; & \NaN & \NaN & \NaN & \NaN & \color{lightgray}0 & \NaN\\
& ST3 & \TableLine 
"tp": 4,
"fn": 0,
"total": 4,
"recall": 1\; & 0.5 & 0.5 & 0 & 0.25 & 1287 & 1 \\ %
& MT3 & \TableLine 
"tp": 8,
"fn": 0,
"total": 8,
"recall": 1\; & 0.125 & 0 & 0.125 & 0 & 671 & 1 \\
& WT3/T4 & \TableLine 
"tp": 23,
"fn": 0,
"total": 23,
"recall": 1\; & 0 & 0 & 0 & 0 & 9213 & 1
}

\def\TableBinarySearch{%
   \AlgoBanner{Binary Search \mtfrmt{57890}}&
T1 & \TableLine 
"tp": 27,
"fn": 89,
"total": 116,
"recall": 0.23275862068965517\; & 1.0 & 1.0 & 0.9655172413793104 & 0.9655172413793104 & \Same116 & \Same0.23275862068965517
\\
& T2 & \TableLine 
"tp": 1,
"fn": 6,
"total": 7,
"recall": 0.14285714285714285\; & 1.0 & 1.0 & 0.7777777777777778 & 1 & \Same7 & \Same0.14285714285714285
\\
& VST3 & \TableLine 
"tp": 7,
"fn": 25,
"total": 32,
"recall": 0.21875\; & 1.0 & 1.0 & 0.875 & 0.9375 & \Same32 & \Same0.21875
\\
& ST3 & \TableLine 
"tp": 111,
"fn": 201,
"total": 312,
"recall": 0.3557692307692308\; & 0.9519230769230769 & 0.9358974358974359 & 0.5384615384615384 & 0.8461538461538461 & \Same312 & \Same0.3557692307692308
\\
& MT3 & \TableLine 
"tp": 655,
"fn": 1451,
"total": 2106,
"recall": 0.31101614434947766\; & 0.008072174738841406 & 0 & 0.04415954415954416 & 0.11253561253561253 & \Same2106 & \Same0.31101614434947766
\\
& WT3/T4 & \TableLine 
"tp": 3292,
"fn": 75136,
"total": 78428,
"recall": 0.041974804916611416\; & 0 & 0 & 0.006757790584995154 & 0.002409853623705819 & \Same78428 & \Same0.041974804916611416
}

\def\TableBubbleSort{\AlgoBanner{Bubble Sort \mtfrmt{23454}}&
T1 & \TableLine 
"tp": 39,
"fn": 0,
"total": 39,
"recall": 1\; & 1.0 & 1.0 & 1 & 1 & \Same39 & \Same1 \\
& T2 & \TableLine 
"tp": 4,
"fn": 0,
"total": 4,
"recall": 1\; & 1.0 & 1.0 & 1 & 1 & \Same4 & \Same1\\
& VST3 & \TableLine 
"tp": 20,
"fn": 1,
"total": 21,
"recall": 0.9523809523809523\; & 0.9523809523809523 & 0.9523809523809523 & 1 & 1 & \Same21 & \Same0.9523809523809523\\
& ST3 & \TableLine 
"tp": 203,
"fn": 4,
"total": 207,
"recall": 0.9806763285024155\; & 0.961352657004831 & 0.961352657004831 & 0.9323671497584541 & 0.3188405797101449 & \Same207 & \Same0.9806763285024155 \\
& MT3 & \TableLine 
"tp": 1576,
"fn": 70,
"total": 1646,
"recall": 0.9574726609963548\; & 0.020656136087484813 & 0.009720534629404616 & 0.255771567436209 & 0.009113001215066828 & \Same1646 & \Same0.9574726609963548 \\
& WT3/T4 & \TableLine 
"tp": 11038,
"fn": 906,
"total": 11944,
"recall": 0.924146014735432\; & 0 & 0 & 0.028717347622237106 & 0 & \Same11944 & \Same0.924146014735432}

\def\TableGreatestCommonDivisor{%
\AlgoBanner{GCD \mtfrmt{2024}}&
T1 &  \TableLine 
"tp": 0,
"fn": 0,
"total": 0,
"recall": null\; & \NaN & \NaN & \NaN & \NaN & 1 & 0 \\
& T2 & \TableLine 
"tp": 0,
"fn": 0,
"total": 0,
"recall": null\; & \NaN & \NaN & \NaN & \NaN & \color{lightgray}0 & \NaN \\
& VST3 & \TableLine 
"tp": 0,
"fn": 0,
"total": 0,
"recall": null\; & \NaN & \NaN & \NaN & \NaN & \color{lightgray}0 & \NaN \\
& ST3 & \TableLine 
"tp": 0,
"fn": 0,
"total": 0,
"recall": null\; & \NaN  & \NaN & \NaN & \NaN & 15 & 0.9333333333333333 \\
& MT3 & \TableLine 
"tp": 0,
"fn": 2,
"total": 2,
"recall": 0\; & 0.0 & 0.0 & 0.0 & 0.0 & 51 & 0.35294117647058826 \\
& WT3/T4 & \TableLine 
"tp": 0,
"fn": 8,
"total": 8,
"recall": 0\; & 0.0 & 0.0 & 0.0 & 0.0 & 186 & 0.3924731182795699
}

\def\TablePalindrome{
\AlgoBanner{Palindrome \mtfrmt{3735}}&
T1 & \TableLine 
"tp": 10878,
"fn": 1,
"total": 10879,
"recall": 0.9999080797867451\; & 1.0 & 1.0 & 1.0 & 1.0 & \Same10879 & \Same0.9999080797867451 \\  %
& T2 & \TableLine 
"tp": 0,
"fn": 0,
"total": 0,
"recall": null\; & \NaN & \NaN & \NaN & \NaN & \color{lightgray}0 & \NaN \\
& VST3 & \TableLine 
"tp": 0,
"fn": 0,
"total": 0,
"recall": null\; & \NaN & \NaN & \NaN & \NaN & \color{lightgray}0 & \NaN \\
& ST3 & \TableLine 
"tp": 149,
"fn": 0,
"total": 149,
"recall": 1\; & 1.0 & 1.0 & 1.0 & 1.0 & \Same149 & \Same1 \\
& MT3 & \TableLine 
"tp": 10,
"fn": 0,
"total": 10,
"recall": 1\; & 0 & 0 & 0.5 & 0.3 & \Same10 & \Same1 \\
& WT3/T4 & \TableLine 
"tp": 1366,
"fn": 1126,
"total": 2492,
"recall": 0.5481540930979133\; & 0 & 0 & 0.48635634028892455 & 0.059791332263242375 & 2990 & 0.4568561872909699%
}

\def\TablePrimeFactors{%
\AlgoBanner{Prime Factors \mtfrmt{1302}}&
T1 & \TableLine 
"tp": 4,
"fn": 0,
"total": 4,
"recall": 1\; & 1 & 1 & 1 & 1 & \Same4 & \Same1 \\
& T2 & \TableLine 
"tp": 3,
"fn": 0,
"total": 3,
"recall": 1\; & 1 & 1 & 1 & 1 & \Same3 & \Same1 \\
& VST3 & \TableLine 
"tp": 3,
"fn": 0,
"total": 3,
"recall": 1\; & 1 & 1 & 1 & 1 & \Same3 & \Same1\\
& ST3 & \TableLine 
"tp": 19,
"fn": 0,
"total": 19,
"recall": 1\; & 1 & 1 & 0.5263157894736842 & 0.8421052631578947 & \Same19 & \Same1 \\
& MT3 & \TableLine 
"tp": 55,
"fn": 5,
"total": 60,
"recall": 0.9166666666666666\; & 0 & 0 & 0.13333333333333333 & 0.1 & \Same60 & \Same0.9166666666666666\\
& WT3/T4 & \TableLine 
"tp": 69,
"fn": 32,
"total": 101,
"recall": 0.6831683168316832\; & 0 & 0 & 0 & 0 & 121 & 0.5702479338842975
}

		\caption{Comparision of our Search Patterns Results per Clone Type as Clone Pairs}
		\label{clone-type-pair-table}
		\def\As{\(\widehat{=}\)}%
		\centering\resizebox*\linewidth!{\begin{tabular}{p{1em}>{\color{gray}}lr@{\hspace{9pt}}S@{\hspace{3pt}}S@{\hspace{3pt}}S@{\hspace{3pt}}S@{\hspace{3pt}}S@{\hspace{12pt}}S[round-mode=none,table-format=5.0]S}
									\hline
												& 				&            \multicolumn{6}{c}{{\large\strut}TPR (Min 50)} & \multicolumn{2}{c}{TPR (Min 0\rlap{)}}\\
										\xcline{3-8} \cline{9-10}
		{}& {\large\strut}& {Total} & {We} & {CW} & {NiCad} & {Oreo} & {SCC} & {Total} & {Ours}\\
				\hline
				\hline
				\TablePrimeFactors\\
				\hline
				\TableGreatestCommonDivisor\\
				\hline
				\TableFibonacci\\
				\hline
				\TablePalindrome\\
				\hline
				\TableBubbleSort\\
				\hline
				\TableBinarySearch\\
				\hline
		\end{tabular}}
	\begin{center}
		\scriptsize CW=CloneWorks, SCC=SourcererCC, We=Our results, (Min~50) and (Min~0) refer to the minimum token limit for BigCloneEval, with identical results greyed out.
	\end{center}
	\end{table}

The purpose of BigCloneEval~\cite{svajlenko2016bigcloneeval} is the evaluation of CCD tools.
	We use it to detect algorithm instances, which makes the comparison with CCD tools difficult.
	The fundamental problem is that we are not interested in finding \textit{pairs} of clones, which is the purpose of the CCD tools.
	To enable a direct comparison, we decided on a simple approach to map our results to the level of clone pairs.
	We iterated through every\footnote{Since the runtime of CCD tools can be quite long BigCloneEval recommends to only consider pairs with a minimal token length of 50~\cite{BigCloneEvalGithub}. 
	We additionally report our results with a minimum token limit of 0, since many \textit{Fibonacci} and \textit{GCD} implementations fall below this limit.} 
	clone pair and counted it as a success if we were able to find both locations with our search pattern.
	If one or both of the locations were not detected by our prototype, the pair is counted as a false negative.
	Using this approach, we calculated the recall of our prototype on a clone pair level, which is comparable to CCD tools.
	To interpret the data it is important to keep in mind that we created our search patterns manually using only reference implementations from the internet.
	The CCD tools on the other hand can extract information from all the implementations in the dataset, which also includes prototypical algorithm implementation examples provided by the BigCloneEval creators.
	However, they have to detect the clones without additional human input.

Since there is no consensus on how to split Type-3 and Type-4 clones BigCloneEval divides them into the following categories (based on their syntactic similarity)~\cite{svajlenko2016bigcloneeval}:
	\begin{description}[widest={WT3/T4:},leftmargin=*]
			\item[\hphantom{A}VST3:] Very-Strongly Type-3\hfill \([\qty{90}{\%}, \qty{100}{\%})\)
			\item[\hphantom{A}ST3:] Strongly Type-3\hfill\([\qty{70}{\%}, \qty{90}{\%})\)
			\item[\hphantom{A}MT3:] Moderately Type-3\hfill\([\qty{50}{\%}, \qty{70}{\%})\)
			\item[\hphantom{A}WT3/T4:] Weakly Type-3 or Type-4\hfill\([\qty{0}{\%}, \qty{50}{\%})\)
	\end{description}
	Our results, separated by clone type, are displayed in \cref{clone-type-pair-table}.

We usually observe an exceptional performance of our prototype on Type-1 and Type-2 often reaching a recall of 1.0.
	More importantly, we can also see that our approach is not limited to Type-1 and Type-2 clones, but also achieves a good recall for the other types.
	Our patterns find many clones in the VST3, ST3 and MT3 categories with often only minimal decreases in recall.
	However, when we move further in the direction of Type-4 clones we notice a drop in performance, which is to be expected.
	Type-4 clones can have entirely different ASTs and their detection is a challenge to this day.

As explained earlier, BigCloneEval is typically not used to compare the precision of CCD tools.
	We can therefore only compare the recall of our tool.
	We compare ourselves with these state-of-the-art CCD tools using the data from~\cite{Saini2018}:
	\textit{CloneWorks} (token-\allowbreak based)~\cite{svajlenko2017cloneworks}, \textit{NiCad} (hybrid token-based)~\cite{cordy2011nicad}, \textit{Oreo} (machine learning based)\cite{Saini2018}, and \textit{SourcererCC} (token-based)\cite{Sajnani2016}.

Aside from \textit{Binary Search} we achieve comparable results for Type-1 and Type-2 clones while outperforming the tools in the more challenging Type-3 and Type-4 categories. %
	The only tool that could detect WT3/T4 clones on more than one algorithm was \textit{Oreo}.
	However, we always report a better recall in this category.
	In fact, we always report better recall than any of the CCD tools from the MT3 category onwards.
	Therefore, another insight from our experiment is that regular CCD tools would be very limited when used for algorithm recognition.
	\AnswerBox{3}{
	  The results clearly show that our search patterns have a significantly higher recall than CCD tools, especially on Type-3 and Type-4 clones.
	  Our evaluation also shows that CCD tools are quite limited when used for algorithm recognition.
	}

\subsection{Threats to validity} %
	Concerning the \textit{construct validity} we compare our manually created search patterns based on reference implementations from the web with the results of CCD tools.
	We argue that while CCD tools have to detect the clones without additional input they can use all the implementations contained in the dataset, which also includes the reference implementation used by Svajlenko et al. to create the benchmark.

With respect to \textit{internal validity}, we rely on a subset of BigCloneEval which was labeled by three judges over 216 hours of manual validation efforts and is widely used with over~90 citations according to IEEE Xplore~\cite{Svajlenko2014, Saini2018, Sajnani2016, Farmahinifarahani2019}.
	BigCloneEval contains many types of clones but our interest lies in recognizing algorithms.
	Therefore, we limited our evaluation to the subset of the benchmark containing algorithms.
	We argue that the selected algorithms are part of the more complex functionalities in BigCloneEval.
	Additionally, the algorithms themselves are quite diverse including searching, sorting, factorization, and recursive algorithms.
	We did not use the implementations included in BigCloneEval to create our search patterns to avoid biased results.

During our evaluation of Codellama 213 of the Java methods in BigCloneEval exceeded our chosen model context window of 4096 tokens and could therefore not be processed by the LLM.
	We considered these methods to be classified as negative by the LLM during the evaluation.
	We argue that these methods only represent 0.002\% of the total methods which is a negligible amount and defaulting to the majority class is a reasonable decision.

The main threat to \textit{external validity} is the fact that BigCloneEval only consists of algorithms implemented within a single method.
	While our approach is capable of expressing complex algorithms consisting of multiple related methods, the evaluation of such search patterns is not possible with this benchmark and is currently considered future work.
	Furthermore, the implementation and evaluation of our prototype target the Java programming language.
	Although our conceptual approach could be adopted for other languages, adjustments will most likely be necessary, especially for languages with different programming paradigms.
	 A positive impact on the external validity stems from the fact that BigCloneEval includes real-world source code taken from open-source projects~\cite{Svajlenko2014}.

\section{Related Work}\label{sec-relatedwork} %

\subsection{Code Search}
In the context of code search, Paul and Prakash present their tool SCRUPLE, which allows users to search source code by expressing queries in a pattern language~\cite{paulFrameworkSourceCode1994}.
The pattern language allows for queries regarding nesting, references to identifiers, or (named) wildcards with constraints on the names and entities to which they get bound.
Under the hood, SCRUPLE transforms the query into a nondeterministic finite state automaton (NFA) ~\cite{hopcroftIntroductionAutomataTheory2001}.
An interpreter then runs the NFA with the attributed syntax tree of the source code as input and records a match if the NFA reaches its final state.
In contrast to our work, Paul and Prakash do not focus on algorithm recognition.
Consequently, SCRUPLE lacks features that we consider essential for searching algorithms.
E.g., SCRUPLE does not provide the ability to formulate multiple queries related via their constraints on bindings.
Additionally, it does not support abstraction features such as our \T{loop()} construct.
However, both topics are discussed as future work.

Atkinson, et al. compare different pattern-matching approaches to create a new tool called TAWK~\cite{atkinsonEffectivePatternMatching2006}. Similar to SCRUPLE~\cite{paulFrameworkSourceCode1994}, it uses an AST-based source code presentation and makes use of state machines to match patterns. 
Upon a successful match a user-defined action specified as C code can be executed with the matched element as input. 
With TAWK patterns can be specified by predefined keywords for AST elements (e.g. function declaration or function call) as well as regular expressions. 
Additionally, search patterns can be connected in a pipes-and-filters-like fashion, so the match of a search pattern is passed to the next. 
With regards to algorithm recognition, TAWK shares the same limitations already mentioned for SCRUPLE.

FaCoY is a code-to-code search engine, which given a code snippet as an input retrieves semantically similar codes from an indexed code base. An important aspect of FaCoY is the query alternation strategy. FaCoY extracts key terms from the input code and uses those in combination with information from StackOverflow (a popular Q\&A site for developers) to generate alternate code queries. This indeed helps FaCoY to find more semantically similar code results than other search engines, especially in the range of Type-3 and Type-4 clones. However, their experiments with BigCloneBench only show a recall of 0.41 and 0.10 for these respective categories.

\subsection{Program Concept Recognition}
Also referred to as Programming Plan Recognition~\cite{Quilici1994} or Cliché Recognition~\cite{Wills1994}, Program Concept Recognition aims at supporting program comprehension by recognizing language-independent ideas of computation and problem-solving methods.
These include general coding strategies, data structures, algorithms, architectural concepts, and domain concepts~\cite{Kozaczynski1992}. %
In general, all of these approaches use manually specified search patterns and parse the source code into an intermediate representation which is then compared with the search pattern.

Kozaczynski et al. represent concepts as ASTs with additional control- and data-flow constraints.
Recognition is a top-down process comparing the structure of the concept and program ASTs and evaluating the constraints.
In order to match a top-level concept they iterate over all the permutations of the sub-concepts evaluating the consistency of bindings.
After recognition, optional transformation rules can be applied to the code.
We believe that our internal Java DSL is a more descriptive way to describe the AST while at the same time offering more expressive power.
For example, we allow matching the children of an AST node in an arbitrary order which is useful for matching independent variable declarations.
Additionally, our internal Java DSL offers a better user experience due to IDE features like code completion.
We match related search patterns sequentially which allows us to reject failed matches earlier and only consider all permutations in the worst case.
Kozaczynski et al. show the practical utility of their tooling with an industry evaluation mainly automating interface changes.
They do not evaluate the accuracy of their approach with respect to recognizing concepts in real-world source code.

Quilici~\cite{Quilici1994} extends Kozaczynski's approach trading the ability to recognize every concept located in the code for improved efficiency.
Among other optimizations, concepts are now represented in a hierarchy with simpler ones triggering the matching of more complex ones resulting in a hybrid bottom-up/top-down search.
An evaluation of these optimizations and the accuracy of the approach is considered future work.

Wills~\cite{Wills1994} represents programs as flow graphs encoding dataflow between functions augmented with control-flow attributes.
Concepts are encoded as graph grammar rules and recognized by parsing the flow graph in accordance with these rules.
Wills highlights the need for future empirical studies focusing on the accuracy and scalability of the approach.
We believe that using our pattern language is a more intuitive formalism for representing concepts compared to graph grammars since many programmers are already familiar with ASTs.

Nunez et al.~\cite{Nunez2017, ARCC2} rely on PDGs and focus on supporting repetitive code comprehension tasks e.g. the grading coding assignments.
Like Kozaczynski et al. they inefficiently check all permutations of search patterns and methods for concepts involving multiple methods.
Furthermore, Nunez et al. offer limited support for user-defined patterns.
They provide no dedicated language for specifying patterns and instead require users to encode the complete PDG manually in XML.
A search pattern expressing the \textit{conditional increment of a variable in a loop} requires 58 lines of XML, while a comparable pattern in our DSL requires 25 lines.
They evaluate their approach by using concept recognition to automatically detect programming errors in a syntactically generated dataset of mutated student submissions.

Metzger et al. also discuss many of the approaches mentioned in this section (\cite{Kozaczynski1992, Quilici1994, Wills1994}) and conclude that they suffer from usability and scalability issues~\cite[pp. 31-50]{Metzger2000}.

\subsection{Algorithm Recognition}
The approaches of Mesnard et al.~\cite{Mesnard2016} and Metzger et al.~\cite{Metzger2000} focus on formally proving the equivalence of two subprograms, which is not possible in the general case~\cite{Metzger2000}. 
Mesnard et al. focus on the detection of algorithms in binary Android code. 
They represent two programs as horn clauses instead of search patterns and try to prove equivalence via semantic preserving transformations.  
If a common representation can be found via the transformations the code fragments are semantically equivalent, otherwise, the search is inconclusive. 
Investigating the feasibility of their approach is considered future work~\cite{Mesnard2016}.

Metzger et al. focus on a formally correct replacement of (nested) loops with library calls to improve performance, e.g. replacing a matrix multiplication with a corresponding call to an optimized math library.
They extract the code in the loops and generate a canonicalized version of the subprogram using semantics-preserving transformations and heuristics.
Afterward, this representation is compared to a database of algorithm patterns which are also stored in canonicalized form.
They analyze the theoretical runtime but do not report data on an evaluation~\cite{Metzger2000}.

Formal methods trade recall for precision since proving the equivalence of two subprograms is only possible under certain restrictions inherent in the design of the respective approach.
We have a different focus and allow the user to specify how strictly the search pattern of the algorithm should be matched and at what point variability is allowed.
This enables the user to find as many implementations as possible if they are willing to tolerate some false positives. 

It is worth mentioning that none of the program comprehension or algorithm recognition approaches discussed so far has evaluated their recognition ability on real source code.
Additionally, except Nunez et al. none of these works is publicly available making a comparison or evaluation of these approaches impossible.

Another class of approaches extracts features from source code and trains a machine learning model to classify algorithms~\cite{Shalaby2017, Long2022, Zhu2011, Taherkhani2010}.
These approaches do not require the specification of explicit search patterns and instead, learn the characteristics separating algorithm implementations from other code.
Simple approaches use characteristics like the type and \# of variables, the type and \# of operands, \#LOCs, and similar characteristics~\cite{Shalaby2017}.
More complex approaches use specialized ML models to automatically extract features from intermediate representations of the source code, such as the control-flow or data-flow graph~\cite{Long2022}.
This representation is then fed to a classifier like SVM, MLP, or RFT for the final classification of the algorithm.
While these approaches often achieve good accuracy on their limited test sets, their evaluation is often conducted with a small number of algorithms.
Additionally, these approaches often include broad categories like recursion, greedy, or trees which makes the classification easier. 
Extending these approaches to detect additional algorithms of interest requires re-training with new data, which is an unreasonable effort for an end-user.
In addition, most approaches do not address how the search would take place on real source code, i.e. whether the search is done on file level or method level and how more complex algorithms with multiple methods in potentially different files can be found.

\section{Conclusions and Future Work}\label{sec-conclusion} %

We conclude that it is possible to detect algorithms with a good recognition performance by specifying patterns on just the abstract syntax tree, without employing additional techniques like control-flow analysis.
Our evaluation on BigCloneEval shows that we were able to generate search patterns with a good balance of precision and recall for all algorithms but \textit{Binary Search}, which was lacking in recall.

We then compare our results with those of Codellama~\cite{roziere2024codellama}, a large language model that has achieved impressive results in a variety of code-related tasks~\cite{roziere2024codellama, liu2024LLMCodeGenEval, silva2023repairllama}.
Codellama's recall is comparable to ours but we significantly outperform the LLM in terms of precision leading to a better \FMeasure-score.
Furthermore, the LLM is considerably slower, with runtimes taking 17 to 376 times longer than our approach.

Since different implementations of an algorithm can be considered clones of each other, we also compare ourselves to CCD tools. %
The calculated recall shows that our manually created search patterns perform far better than CCD tools.
Especially on Type-3 and Type-4 clones where we generally report better results.
Another conclusion is that CCD tools are not well suited for algorithm recognition due to their poor performance in these categories.

\label{section:future-work}
We have identified multiple opportunities for improvement and initial ideas on how these could be addressed.
Our evaluation has shown that very permissive search patterns can lead to large runtimes when analyzing long and complex methods.
	 This is the case because we currently create a separate matching state for each possible combination of bindings resulting in exponentially many in the worst case.
	 To alleviate this problem we could instead store multiple valid options for each binding in the matching state.
	 Essentially replacing multiple matching states with a set of options, which would improve both runtime and memory usage.

Other ideas to increase performance and precision include:
	 \begin{itemize}
	\item Using heuristics to exclude program ASTs from the matching if they do not meet a set of minimal requirements indicating that a match is even possible. %
	\item Optimizing the matching order of multiple search patterns related via bindings.
	\item Normalizing the AST to reduce variability.
	\item Including data- and control-flow analysis.
	 \end{itemize}
	 Our evaluation shows that our approach can be used to create good search patterns.
	 Nonetheless, further improving the usability of our tooling will be necessary in the future, to enable users to describe search patterns effectively.
	 Ideally, the tooling would allow it to highlight key features of a user-provided reference implementation and then generate a search pattern automatically, offering different levels of abstraction.

A similar research direction is the automatic generation of search patterns from (multiple) reference implementations of an algorithm.
	 Our initial experiments in this direction show that by selecting the right abstraction rules, it is indeed possible to create search patterns with good recall and precision.
	 However, more research is needed on how to effectively and efficiently select or learn suitable abstraction rules, especially considering the large search space of possible options.

Lastly, while our initial results are promising a further evaluation with more complex algorithms is necessary to validate these results.

\begin{anonsuppress}
\section{Acknowledgments}
	The authors acknowledge support by the state of Baden-Württemberg through bwHPC.
\end{anonsuppress}

\bibliographystyle{IEEEtran}
\bibliography{references}

\end{document}